\documentclass[fleqn,usenatbib]{mnras}

\usepackage{newtxtext,newtxmath}

\usepackage[T1]{fontenc}

\DeclareRobustCommand{\VAN}[3]{#2}
\let\VANthebibliography\thebibliography
\def\thebibliography{\DeclareRobustCommand{\VAN}[3]{##3}\VANthebibliography}

\usepackage{graphicx}	
\usepackage{amsmath}	

\hypersetup{linkcolor=magenta,citecolor=cyan,filecolor=green,urlcolor=blue}

\usepackage{array}
\usepackage{booktabs}
\usepackage{float}
\usepackage{threeparttable}
\usepackage{tabularx}
\usepackage{enumitem}
\usepackage{multirow}
\usepackage[]{xcolor}

\title[ADAF for state transition in BH-XRBs]{Advection-dominated accretion flow for the varied transition luminosities in black hole X-ray binaries}

\author[J. Li and E. Qiao]{
Jiaqi Li,$^{1,2}$
Erlin Qiao,$^{1,2}$\thanks{E-mail: jqli@bao.ac.cn, qiaoel@nao.cas.cn}
\\
$^{1}$National Astronomical Observatories, Chinese Academy of Sciences, 20A Datun Road, Beijing 100101, China\\
$^{2}$School of Astronomy and Space Sciences, University of Chinese Academy of Sciences, 19A Yuquan Road, Beijing 100049, China\\
}

\date{Accepted XXX. Received YYY; in original form ZZZ}

\pubyear{2023}

\begin{document}
\label{firstpage}
\pagerange{\pageref{firstpage}--\pageref{lastpage}}
\maketitle

\begin{abstract}
Observationally, two main spectral states, i.e., the low/hard state and the high/soft state, are identified in black hole X-ray binaries (BH-XRBs). 
Meanwhile, the transitions between the two states are often observed. 
In this paper, we re-investigate the transition luminosities in the framework of the self-similar solution of the advection-dominated accretion flow (ADAF). 
Specifically, we search for the critical mass accretion rate $\dot m_{\rm crit}$ of ADAF for different radii $r$ respectively. It is found that $\dot m_{\rm crit}$ decreases with decreasing $r$. By testing the effects of BH mass $m$, the magnetic parameter $\beta$ and the viscosity parameter $\alpha$, it is found that only $\alpha$ has significant effects on $\dot m_{\rm crit}-r$ relation.
We define the minimum $\dot m_{\rm crit}$ (roughly at the innermost stable circular orbit) as the hard-to-soft transition rate $\dot m_{\rm tr: H\rightarrow S}$, above which BH will gradually transit from the low/hard state to the high/soft state, and $\dot m_{\rm crit}$ at $30$ Schwarzschild radii as the soft-to-hard transition rate $\dot m_{\rm tr: S\rightarrow H}$, below which BH will gradually transit from the high/soft state to the low/hard state. 
We derive fitting formulae of $\dot m_{\rm tr: H\rightarrow S}$ and $\dot m_{\rm tr: S\rightarrow H}$ as functions of $\alpha$ respectively.
By comparing with observations, it is found that the mean value of $\alpha$ are $\alpha \sim 0.85$ and $\alpha \sim 0.33$ for the hard-to-soft transition and the soft-to-hard transition respectively, which indicates that two classes of $\alpha$ are needed for explaining the hysteresis effect during the state transition. Finally, we argue that such a constrained $\alpha$ may provide valuable clues for further exploring the accretion physics in BH-XRBs.

\end{abstract}


\section{introduction}\label{sec1}
Black hole X-ray binaries (BH-XRBs) are binary systems that contain a primary star of a BH and a secondary star (\citealt{remillard2006,done2007,liu2007}).
Observationally, two main spectral states, i.e., the low/hard state and the high/soft state, are identified in BH-XRBs (\citealt{remillard2006,belloni2010}). 
In the low/hard state (usually with low luminosities), BH-XRBs typically show the hard X-ray spectra dominated by a power-law component with a photon spectral index of about $1.8$ and a cut-off at a few hundred keV, which is often explained by the advection-dominated accretion flow (ADAF)(\citealt{ichimaru1977,narayan1994,narayan1995a,narayan1995b,abramowicz1995,chen1995,esin1997,mahadevan1997,manmoto1997,yuan2014} for review). 
In the high/soft state (usually with high luminosities), their spectra are dominated by a thermal component with a temperature of about $1$ keV, which is often explained by the standard accretion disk expanding to the innermost stable circular orbit (ISCO) of a BH (\citealt{shakura1973}). 

The transition between the low/hard state and the high/soft state (and vice versa) is often observed (\citealt{tanaka1996,fender1999,homan2001,maccarone2003a}). 
\cite{tananbaum1972} firstly observed the state transition in Cyg X-1. 
\cite{miyamoto1995} calculated the ratio of the luminosities of a thermal/power-law component to the total luminosities to determine the time that the state transition occurs. 
\cite{maccarone2003a} collected a sample composed of ten X-ray binaries, showing that the luminosities of the transition from the high/soft state to the low/hard state are about $1-4 \%$ of the Eddington luminosity. 
In addition, the hysteresis effect, i.e., the transition luminosity from the low/hard state to the high/soft state roughly $3-5$ times higher than that of from the high/soft state to the low/hard state, is reported in several sources during the state transition (\citealt{nowak2002,maccarone2003b,zdziarski2004,meyer2005}).
Furthermore, the relation between the state transition luminosities and other factors, e.g. the mass accretion rate, the hardness ratio (HR) of different bands, the inclination angle of sources, etc., were investigated (\citealt{belloni2010,dunn2010,motlagh2019}). 
  
The observed state transition phenomenon and the hysteresis effect have been studied for many years (\citealt{esin1997,dubus2001,meyer2000a,meyer2000b,meyer2005}), but there is no definite conclusion about its physical origin. 
The disk evaporation model is one of the most promising models for the state transition, in which the evaporation rate increases with decreasing radius until a maximum evaporation rate is reached at a few tens or a few hundred of the Schwarzschild radii (\citealt{meyer2000b,meyer2000a,liu2002}). 
If the initial mass accretion rate in the disk is less than the maximum evaporation rate, the disk will truncate at a radius where the mass accretion rate equals the evaporation rate. 
Less than this truncation radius, the accretion flow will be in the form of ADAF, which can well explain the spectral features of the low/hard state in BH-XRBs.
With the increase of the mass accretion rate, if the initial mass accretion rate is larger than the maximum evaporation rate, the disk will extend down to the ISCO of the BH. 
In this case, the emission is dominated by the disk, and the BH-XRBs transit to the high/soft state. 
This maximum evaporation rate represents the critical mass accretion rate for the hard-to-soft transition, and it is found that this critical mass accretion rate is very sensitive to the viscosity parameter $\alpha$, as $\dot m_{\rm crit} \propto \alpha^{2.34}$ ( with $\dot m_{\rm crit}$ scaled with the Eddington accretion rate) (\citealt{qiao2009}).
Based on the work of \citet{meyer2000b,meyer2000a} and \cite{liu2002}, in the paper of \cite{meyer2005}, the authors further studied a case in which initially the disk extends down to ISCO of the BH, corresponding to the high/soft state. 
In this case, the authors re-studied the evaporation process of the accretion disk, and it is found that the whole evaporation rate along the radii is suppressed due to the strong Compton cooling of the soft photon from the inner disk to the outer corona.
Specifically, the maximum evaporation rate decreases by a factor of 3-5 times depending on the model parameters. 
This theoretical result is then readily applied to explain the lower luminosities of the soft-to-hard transitions, i.e., the hysteresis effect observed in BH-XRBs (\citealt{meyer2005,liubf2005}). 

Several other models have also been proposed for explaining the state transition in BH-XRBs.
The generation and transport of magnetic field may be one of the factors for the state transition (\citealt{petrucci2008,latter2012,yan2015,begelman2014,begelman2015}). 
Based on the magnetically arrested discs (MADs)(\citealt{bisnovatyi-kogan1974,narayan2003}), large-scale magnetic reconnection provides an alternative mechanism for regulating the mass accretion rate (\citealt{dexter2014,mckinney2012}). 
\cite{caoxw2016} suggested the model of ADAF with magnetically driven outflow to explain the hysteresis effect during the state transition of BH-XRBs. 
In the above works, the mechanisms may be different, but the state transition and the hysteresis effect are generally believed to be directly related to the mass accretion rate. 

The study of the ADAF solution itself from the point of energy balance also predicts the critical mass accretion rate $\dot m_{\rm crit}$, above which the ADAF solution can not exist and is replaced by the standard accretion disk (\citealt{narayan1994,narayan1995b}). 
This critical mass accretion rate represents a mass accretion rate for the spectral state transition between the low/hard state and the high/soft state. 
Rough estimates with the self-similar solution of ADAF show that this critical mass accretion rate is dependent on the viscosity parameter $\alpha$, specifically expressed as $\dot m_{\rm crit} \propto \alpha^{2}$ (\citealt{mahadevan1997}), which is roughly consistent with the results of the disk evaporation model for the hard-to-soft transition (\citealt{qiao2009,taam2012}). 
In \cite{esin1997}, the authors calculated $\dot m_{\rm crit}$ as a function of $r$ with the global solution of ADAF. 
In the calculations, the authors always take the advection factor $f=0.35$ ($f$ describing the advected fraction of the viscously dissipated energy), which however does not always hold as will be demonstrated in Section 3 of this paper.

In this paper, we re-investigate the transition luminosities of BH-XRBs in the framework of the self-similar solution of ADAF. Compared with previous studies, we search for $\dot m_{\rm crit}$ for different $r$ separately by self-consistently calculating the structure of ADAF ranging from ISCO of a non-rotating BH to $1000 R_{\rm S}$.  
We test the effects of BH mass $m$, the magnetic parameter $\beta$, and the viscosity parameter $\alpha$ on the relation between $\dot m_{\rm crit}$ and $r$. 
It is found that the effects of $m$ and $\beta$ on the relation between $\dot m_{\rm crit}$ and $r$ are very weak and nearly can be neglected. 
 Meanwhile, it is found that the relation between $\dot m_{\rm crit}$ and $r$ can be strongly affected by changing $\alpha$.
We define the minimum $\dot m_{\rm crit}$ as the hard-to-soft transition rate $\dot m_{\rm tr: H\rightarrow S}$, and define $\dot m_{\rm crit}$ at $30$ Schwarzschild radii as the soft-to-hard transition rate $\dot m_{\rm tr: S\rightarrow H}$.
We derive fitting formulae of $\dot m_{\rm tr: H\rightarrow S}$ and $\dot m_{\rm tr: S\rightarrow H}$ as functions of $\alpha$.
Further, by comparing with the observed transition luminosities of both the hard-to-soft transition and the soft-to-hard transition, we constrain the values of $\alpha$ of the hard-to-soft transition and the soft-to-hard transition respectively. 
Finally, we discuss the potential physical meanings for the constrained values of $\alpha$.
The model is briefly introduced in Section \ref{sec2}. 
The numerical results and the comparison with observation are shown in Sections \ref{sec3} and \ref{sec4}. 
The discussion and conclusion are given in Sections \ref{sec5} and \ref{sec6}.

\section{Model}\label{sec2}
In this paper, we search for the critical mass accretion rate $\dot m_{\rm crit}$ of the self-similar solution of ADAF for different model parameters. 
For clarity, we list the equations for ADAF as follows (\citealt{narayan1995b}).

Equation of state,
\begin{equation}\label{EOS}
p_{\rm g}=\beta \rho c_{\rm s}^{2}={ {\rho k T_{\rm i}}\over {\mu_{\rm i} m_{\rm p}} }
+{ {\rho k T_{\rm e}} \over {\mu_{\rm e}m_{\rm p}} }, 
\end{equation}
where $p_{\rm g}$ is the gas pressure, $\beta$ is the magnetic parameter (defined as $\beta=p_{\rm g}/p$, with $p=p_{\rm g}+p_{\rm m}$, and $p_{\rm m}=B^2/{8\pi}$ being the magnetic pressure), $m_{\rm p}$ is the proton mass, $T_{\rm{i}}$ is the ion temperature, $T_{\rm{e}}$ is the electron temperature, $\mu_{\rm{i}}$ and $\mu_{\rm{e}}$ are the effective molecular weights of ions and electrons,
\begin{equation}\label{MW}
\begin{aligned}
&\mu_{\rm{i}}=\frac{4}{1+3X}=1.23,\\
&\mu_{\rm{e}}=\frac{2}{1+X}=1.14.
\end{aligned}
\end{equation}
The hydrogen mass fraction $X=0.75$ is adopted for $\mu_{\rm{i}}$ and $\mu_{\rm{e}}$.

We list the basic physical quantities derived from the self-similar solution of ADAF as,
\begin{equation}\label{SSS}
\begin{aligned}
&v =  2.12 \times 10^{10} \alpha c_{1} r^{-1/2} \  \  \rm {cm\;s^{-1}},\\
&\Omega = 7.19 \times 10^{4} c_{2} m^{-1} r^{-3/2} \ \  \rm{s^{-1}},\\
&{c_{\rm{s}}}^{2} = 4.50 \times  10^{20} c_{3} r^{-1} \ \ \rm{cm\;s^{-2}},\\
&\rho = 3.79 \times 10^{-5} \alpha^{-1} c_{1}^{-1} c_{3}^{1/2} m^{-1} \dot{m} r^{-3/2}\ \  \rm{g\;cm^{-1}},\\
&p = 1.71 \times 10^{16} \alpha^{-1} c_{1}^{-1} c_{3}^{1/2} m^{-1} \dot{m} r^{-5/2}\ \ \rm{g\;cm^{-3}\;s^{-2}},\\
&B = 6.55 \times 10^{8} \alpha^{-1/2} (1-\beta)^{1/2} c_{1}^{-1/2} c_{3}^{1/4} m^{-1/2} \dot{m}^{1/2} r^{-5/4}\ \ \rm{G},\\
&q^{+} = 1.84 \times 10^{21} \epsilon' c_{3}^{1/2} m^{-2} \dot{m} r^{-4}\ \  \rm{erg\;cm^{-3}\;s^{-1}},\\ 
&n_{\rm{e}} = 2.0 \times 10^{19} \alpha^{-1} c_{1}^{-1}  c_{3}^{-1/2} m^{-1} \dot{m} r^{-3/2}\  \ \rm{cm^{-3}},\\
&\tau_{\rm{es}} = 12.4 \alpha^{-1} c_{1}^{-1} \dot{m} r^{-1/2}\ \ \rm{cm^{-3}},
\end{aligned}
\end{equation}
where $v$ is the radial velocity, $\Omega$ is the angular velocity, $c_{\rm{s}}$ is the isothermal sound speed, $\rho$ is the density, $p$ is the pressure, $B$ is the magnetic field, $n_{\rm{e}}$ is the electron number density, $q^{+}$ is the viscous dissipation of energy per unit volume, which are all expressed as functions of $m$, $\dot{m}$, $r$, $\alpha$ and $\beta$.
Here $\alpha$ is the viscosity parameter, $m$ is the BH mass in units of the solar mass $M_{\odot}$, $r$ is the radius in units of the Schwarzschild radius $R_{\rm S}$ (with $R_{\rm S} = \frac{2GM}{c^{2}} = 2.95 \times 10^{10}m \ {\rm cm}$), $\dot m$ is the mass accretion rate in units of the Eddington accretion rate 
$\dot M_{\rm Edd}$ (with $\dot{M}_{\rm{Edd}} = \frac{L_{\rm{Edd}}}{\eta c^{2}} = \frac{4\pi G M}{\eta \kappa_{\rm{es}} c} = 1.39 \times 10^{18} m\ \rm{g \; s^{-1} }$, 
$L_{\rm Edd}$ being the Eddington luminosity, $\eta$ being the radiative efficiency and $\eta=0.1$ adopted), and 
\begin{equation}\label{CFT}
\begin{aligned}
&c_{1} = \frac{3}{5+2 \epsilon'}, \\
&c_{3} = \frac{2}{5+2 \epsilon'},\\
&\epsilon' = \frac{\epsilon}{f}, \\
&\epsilon = (\frac{5/3-\gamma}{\gamma-1}),\\
&\gamma = \frac{9-3\beta}{6-3\beta},
\end{aligned}
\end{equation}
with $f$ being the advected fraction of the viscously dissipated energy. 
Substituting $\rho$ and $c_{\rm{s}}^{2}$ into equation (\ref{EOS}), the equation of state of gas can be rewritten as,
\begin{equation}\label{EQSTATE}
\begin{aligned}
&T_{\rm{i}} + 1.08T_{\rm{e}} = 6.66 \times 10^{12} \beta c_{\rm{3}}r^{-1}.
\end{aligned}
\end{equation}
The energy equations can be expressed as,
\begin{align}
&q^{+} = fq^{+} + q^{\rm{ie}},\label{EQEION}\\ 
&q^{\rm{ie}} = q^{-}, \label{EQEELC}
\end{align}
where $q_{\rm{ie}}$ is the energy transfer rate from ions to electrons via Coulomb collision (\citealt{stepney1983}), which can be expressed as,
\begin{equation}\label{eq11}
\begin{aligned}
&q_{\rm{ie}} = 3.59 \times 10^{-32} n_{\rm{e}}n_{\rm{i}}(T_{\rm{i}}-T_{\rm{e}}) \frac{1+T^{'1/2}}{T^{'3/2}},
\end{aligned}
\end{equation}
with $T^{'} = \frac{kT_{\rm{e}}}{m_{\rm{e}}c^{2}}(1+\frac{m_{\rm{e}}}{m_{\rm{p}}}\frac{T_{\rm{i}}}{T_{\rm{e}}})$.
$q^{-}$ is the electron cooling rate with 
$q^{-}=q_{\rm{brem}}^{-}+q_{\rm{syn}}^{-}+q_{\rm{brem,C}}^{-}+q_{\rm{syn,C}}^{-}$. 
$q_{\rm{brem}}^{-}, q_{\rm{syn}}^{-}, q_{\rm{brem,C}}^{-}$, and $q_{\rm{syn,C}}^{-}$ are the bremsstrahlung cooling rate, the synchrotron cooling rate, the self-Comptonization cooling rate of bremsstrahlung radiation, and the self-Comptonization cooling rate of synchrotron radiation, respectively. 
The specific expression of $q_{\rm{brem}}^{-}, q_{\rm{syn}}^{-}, q_{\rm{brem, C}}^{-}, q_{\rm{syn, C}}^{-}$ can be referred to \cite{narayan1995b}, which are all functions of the advected fraction of the viscously dissipated energy $f$, the ion temperature $T_{\rm i}$ and the electron temperature $T_{\rm e}$.

We solve the equations (\ref{EQSTATE}), (\ref{EQEION}) and (\ref{EQEELC}) for $T_{\rm i}$,  $T_{\rm e}$ and $f$ by specifying $m$, $\dot m$, $r$, $\alpha$ and $\beta$.
As discussed in several previous papers, there should be a critical value mass accretion rate, i.e., $\dot m_{\rm crit}$, only below which the solution of equations (\ref{EQSTATE}), (\ref{EQEION}) and (\ref{EQEELC}) can exist (\citealt{narayan1995b}).
In the following section, we search for $\dot m_{\rm crit}$ for different model parameters, i.e., $m$, $r$, $\alpha$ and $\beta$ of ADAF. 

\section{Numerical Results}\label{sec3}
\subsection{$\dot{m}_{\rm{crit}}$ as a function of $r$ }\label{sec31}
\begin{figure*}
	\centering
        \includegraphics[width=0.47\textwidth]{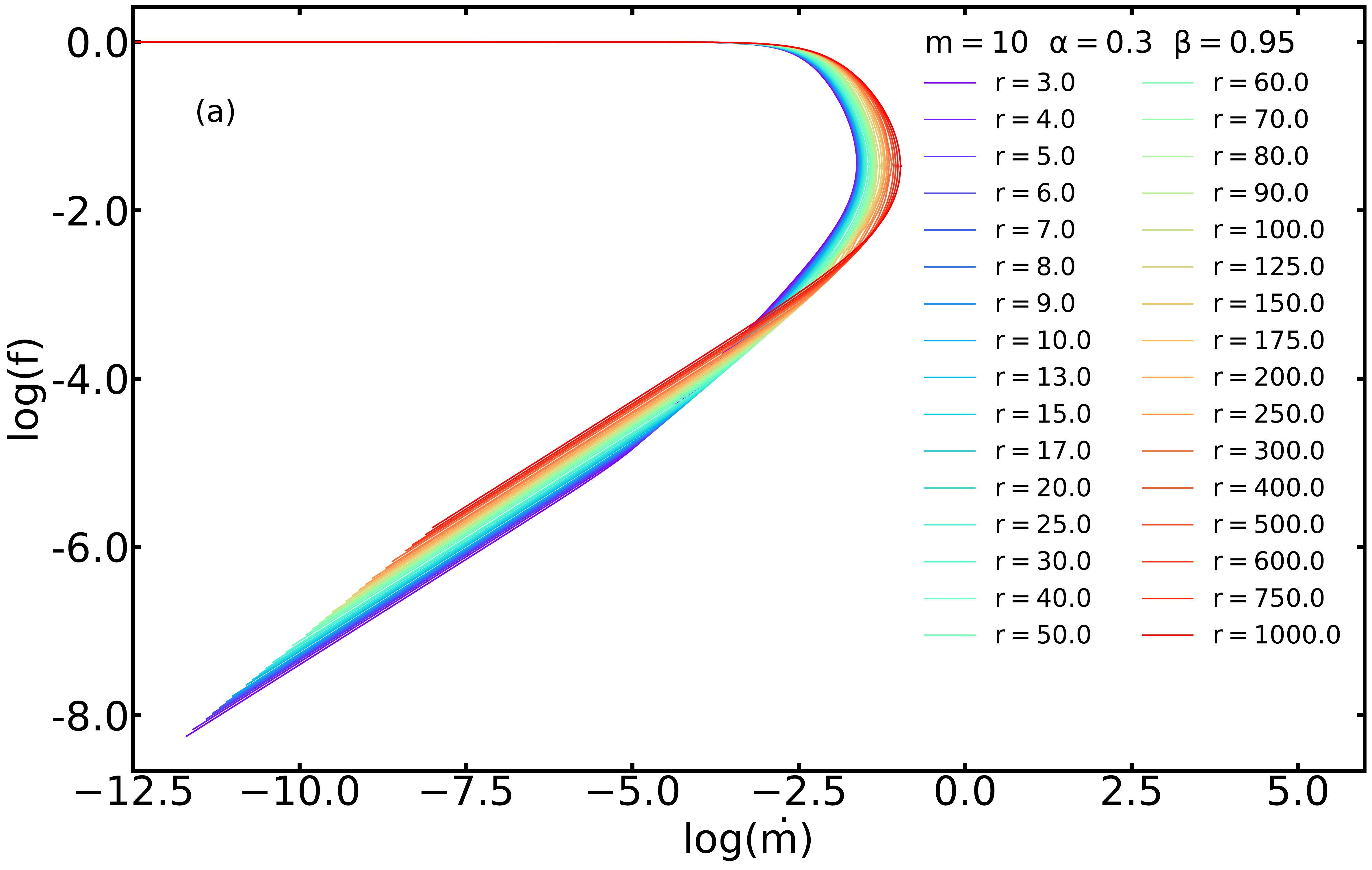}
        \includegraphics[width=0.47\textwidth]{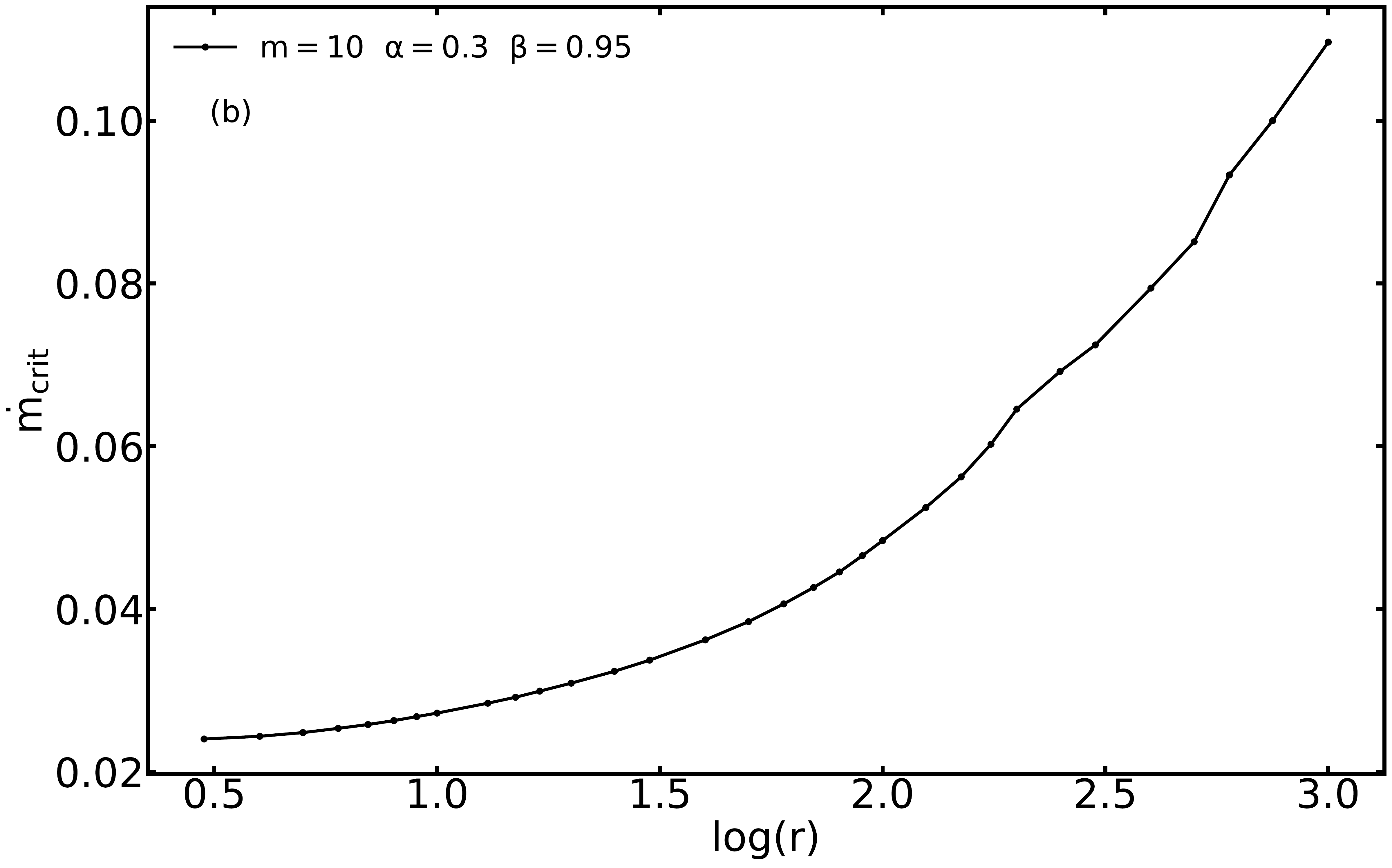}
	\vspace{-0.3cm}
    \caption{Panel (a): The advected fraction of the viscously dissipated energy $f$ as a function of $\dot{m}$ for different $r$ with $m=10$, $\alpha=0.3$, and $\beta=0.95$. 
    Panel (b): The critical accretion rate $\dot m_{\rm crit}$ as a function of $r$ with $m=10$, $\alpha=0.3$, and $\beta=0.95$.}
    \label{mdotcrit}
\end{figure*}

In panel (a) of Figure \ref{mdotcrit}, we plot the advected fraction of the viscously dissipated energy $f$ as a function of $\dot{m}$ for different $r$ with $m=10$, $\alpha=0.3$ and $\beta=0.95$.  
Given a fixed radius, clearly, it is found that there is a vertex-like point. 
We define $\dot m$ at this point as $\dot m_{\rm crit}$, above which there is no solution of the equation system of (\ref{EQSTATE}),  (\ref{EQEION}) and (\ref{EQEELC}).
Below $\dot m_{\rm crit}$, there are two branches for the relation between $f$ and $\dot m$. 
\footnote{Actually, there are three branches below $\dot m_{\rm crit}$, i.e., the uppermost ADAF branch, the lowest cooling-dominated branch, and the middle unstable branch (\citealt{narayan1995b}). Since in the present paper, we focus on the search for $\dot m_{\rm crit}$, we do not plot the lowest cooling-dominated branch for simplicity.} 
The upper branch corresponds to the solution of the advection-dominated accretion flow, i.e., ADAF, which is the solution we focus on in the present paper. 
ADAF is a kind of hot accretion flow with the ion temperature $T_{\rm i}$ being very close to the virial temperature and the electron temperature $T_{\rm e}\sim 10^9 K$. 
In panel (a) of Figure \ref{mdotcrit}, we can see that, in a wide range of $\dot m$ below $\dot m_{\rm crit}$, $f$ is very close to unity, which is the essence of ADAF solution, i.e., most of the viscously dissipated energy is advected in the event horizon of a BH.
As $\dot m$ approaches to $\dot m_{\rm crit}$, $f$ decreases. 
It is found that for nearly all radii, $f$ is $\sim 0.035$ at $\dot m_{\rm crit}$.
The ADAF solution is both thermally stable and viscously stable (\citealt {narayan1995b}). 
The lower branch is unstable, which actually corresponds to the SLE solution of \cite{shapiro1973}.

\begin{figure}
	\centering
    \includegraphics[width=0.47\textwidth]{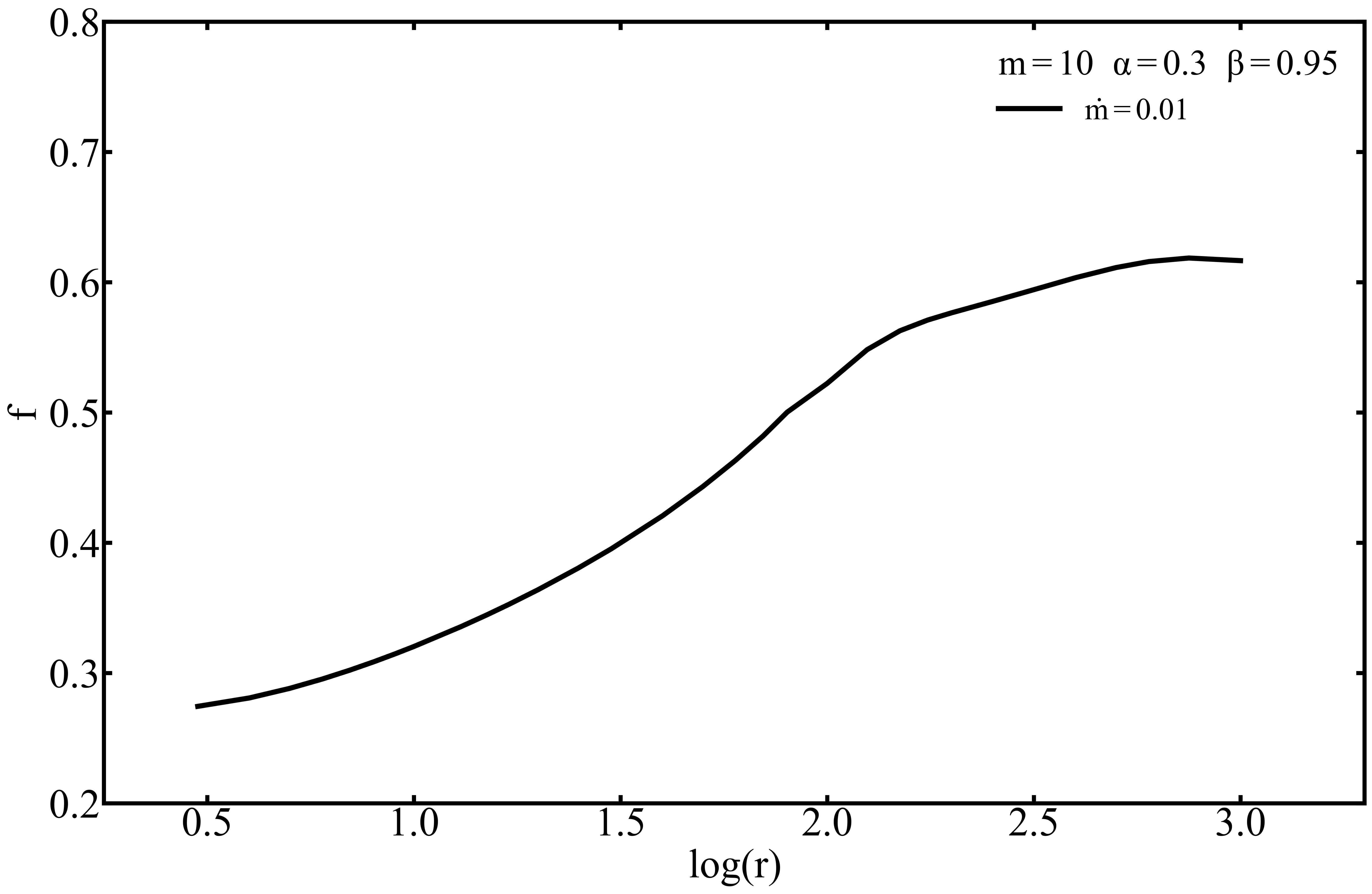}
	\vspace{-0.3cm}
	\caption{The advected fraction of the viscously dissipated energy $f$ as a function of $r$ with $m=10$, $\alpha=0.3$, and $\beta=0.95$ for a fixed mass accretion rate of $\dot{m} = 0.01$.} 
	\label{randf}
\end{figure}

Based on the derived $\dot m_{\rm crit}$ for different $r$, in panel (b) of Figure \ref{mdotcrit}, we plot $\dot m_{\rm crit}$ as a function of $r$. 
It is found that $\dot m_{\rm crit}$ decreases with decreasing $r$.
Specifically, for $r=1000$, the critical mass accretion rate is $\dot m_{\rm crit}=0.1$, and for $r=3$, the critical mass accretion rate is $\dot m_{\rm crit}=0.024$. 
The relation between $\dot m_{\rm crit}$ and $r$ can be understood as follows.   
According to the equation (\ref{EQEION}), the value of $f$ can be expressed as $f=1-q_{\rm ie}/q^{+}$. 
Here as an example, in Figure \ref{randf}, we plot $f$ as a function of $r$ for a fixed mass accretion rate of $\dot m=0.01$ as taking $m=10$, $\alpha=0.3$, and $\beta=0.95$. 
It is clear that $f$ decreases with decreasing $r$, which reversely means that the radiative fraction of the viscously dissipated energy, i.e., $1-f$, increases with decreasing $r$. 
It is easy to imagine that as $1-f$ increases, the mass accretion rate, i.e., $\dot m=0.01$, is closer to $\dot m_{\rm crit}$ locally as discussed in \citep{narayan1994,narayan1995b} and also in the first paragraph of Section \ref{sec31} of this paper. So the increase of $1-f$ (or the decrease of $f$) with decreasing $r$ intrinsically means that $\dot m_{\rm crit}$ decreases with decreasing $r$. 
We note that a relation of $\dot m_{\rm crit} \propto r^{-1/2}$
is predicted in \cite{abramowicz1995}, which is different from the results shown in this paper. The difference of the relation of $\dot m_{\rm crit}$ and
$r$ between \cite{abramowicz1995} and our results in this paper may be caused by the different considerations of the dynamics of the accretion flow, e.g., the global solution adopted in \cite{abramowicz1995} and the self-similar solution adopted in our paper, the only simple assumption of Keplerian motion for the angular velocity of ADAF in \cite{abramowicz1995} and the self-consistent calculation of the  angular velocity of ADAF in our paper (the angular velocity of ADAF actually to be sub-Keplerian in nature) and so on. 

\subsection{The effect of BH mass $m$ on $\dot m_{\rm crit}$} \label{sec32}
\begin{figure}
	\centering
    \includegraphics[width=0.47\textwidth]{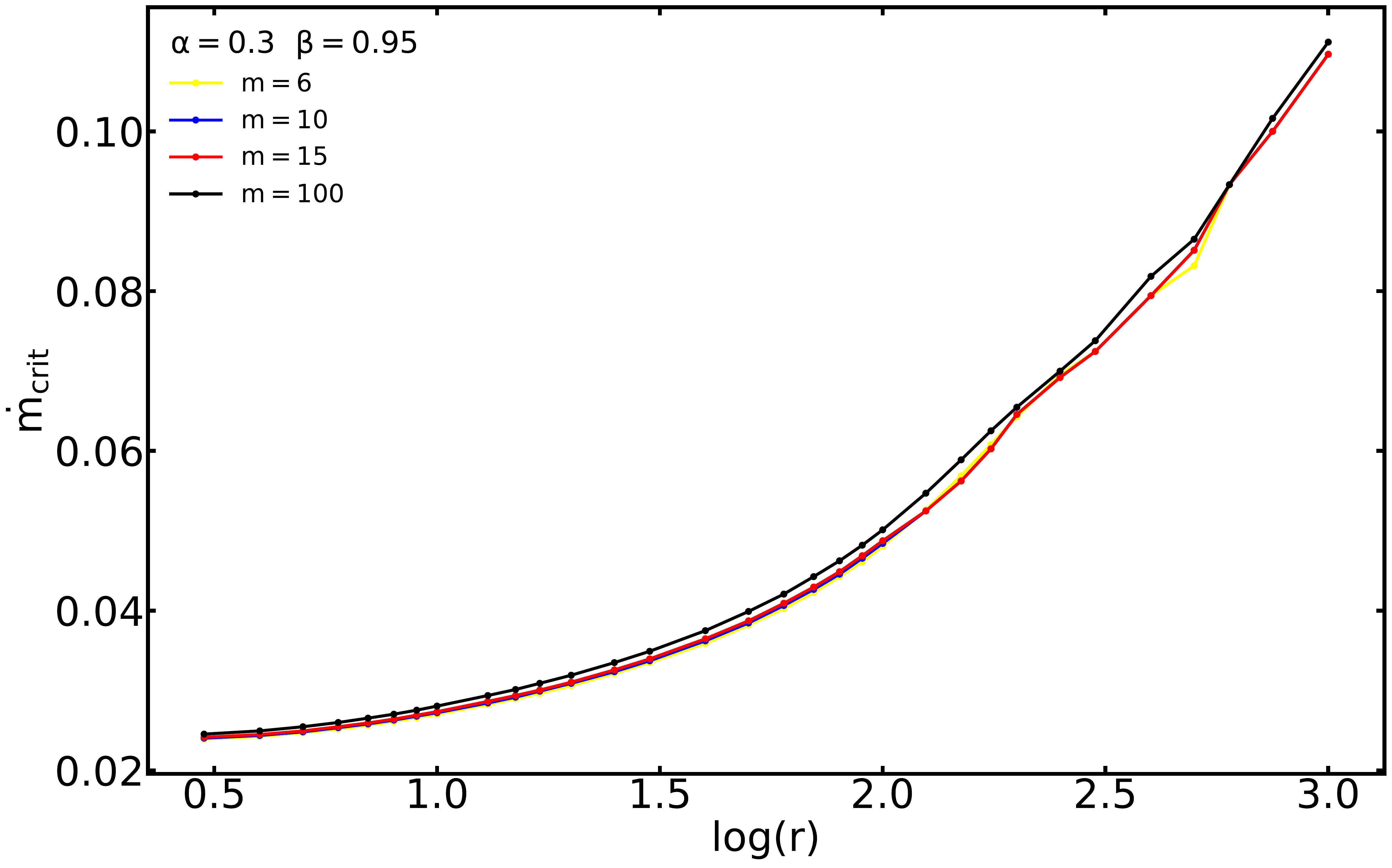}
	\vspace{-0.3cm}
	\caption{The critical mass accretion rate $\dot m_{\rm crit}$ as a function of $r$ for different $m$ with $\alpha=0.3$, and $\beta=0.95$.} 
	\label{theeffectofm}
\end{figure}

In Figure \ref{theeffectofm}, we plot $\dot m_{\rm crit}$ as a function of $r$ for different BH masses of $m=6, 10, 15, 100$ with $\alpha=0.3$, $\beta=0.95$. 
As a whole, it is found that the effect of $m$ on $\dot m_{\rm crit}$ is very weak, and nearly can be neglected. 
Our calculations confirm the predictions of the weak dependence of $\dot m_{\rm crit}$ on $m$ in the ADAF solution (\citealt{narayan1995b,mahadevan1997}). 
Comparing with the previous calculations, e.g. \cite{narayan1995b} and \cite{mahadevan1997}, we search for $\dot m_{\rm crit}$ one by one for different $r$ rather than for $\dot m_{\rm crit}$ with very roughly analysis from the point of energy balance. Specifically, our finding is that the effect of $m$ on $\dot m_{\rm crit}$ is very weak in every $r$ ranging from $r=3$ to $r=1000$.
		 
\subsection{The effect of the magnetic parameter $\beta$ on $\dot m_{\rm crit}$} \label{sec33}
\begin{figure*}
	\centering
    \includegraphics[width=0.47\textwidth]{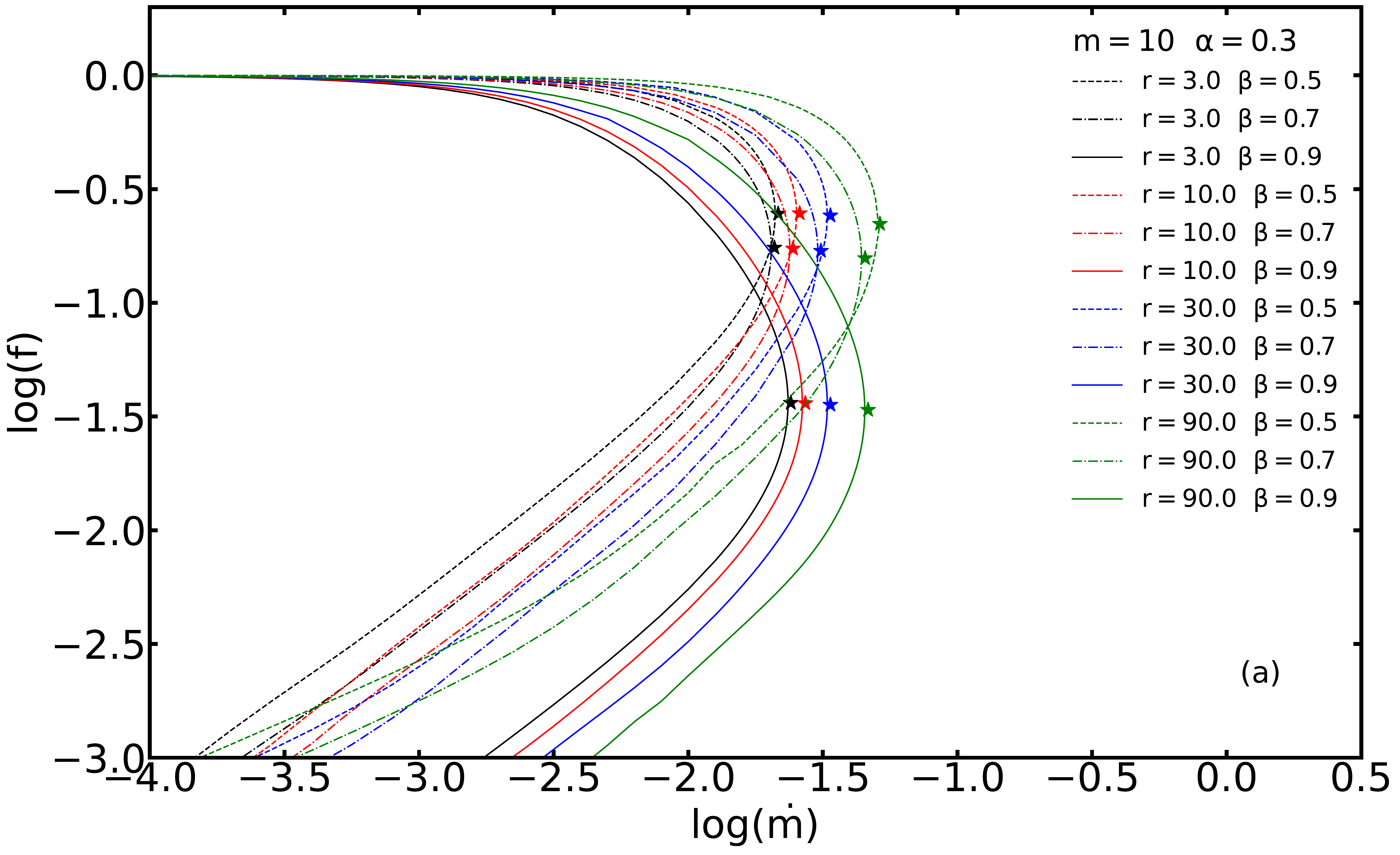}
    \includegraphics[width=0.47\textwidth]{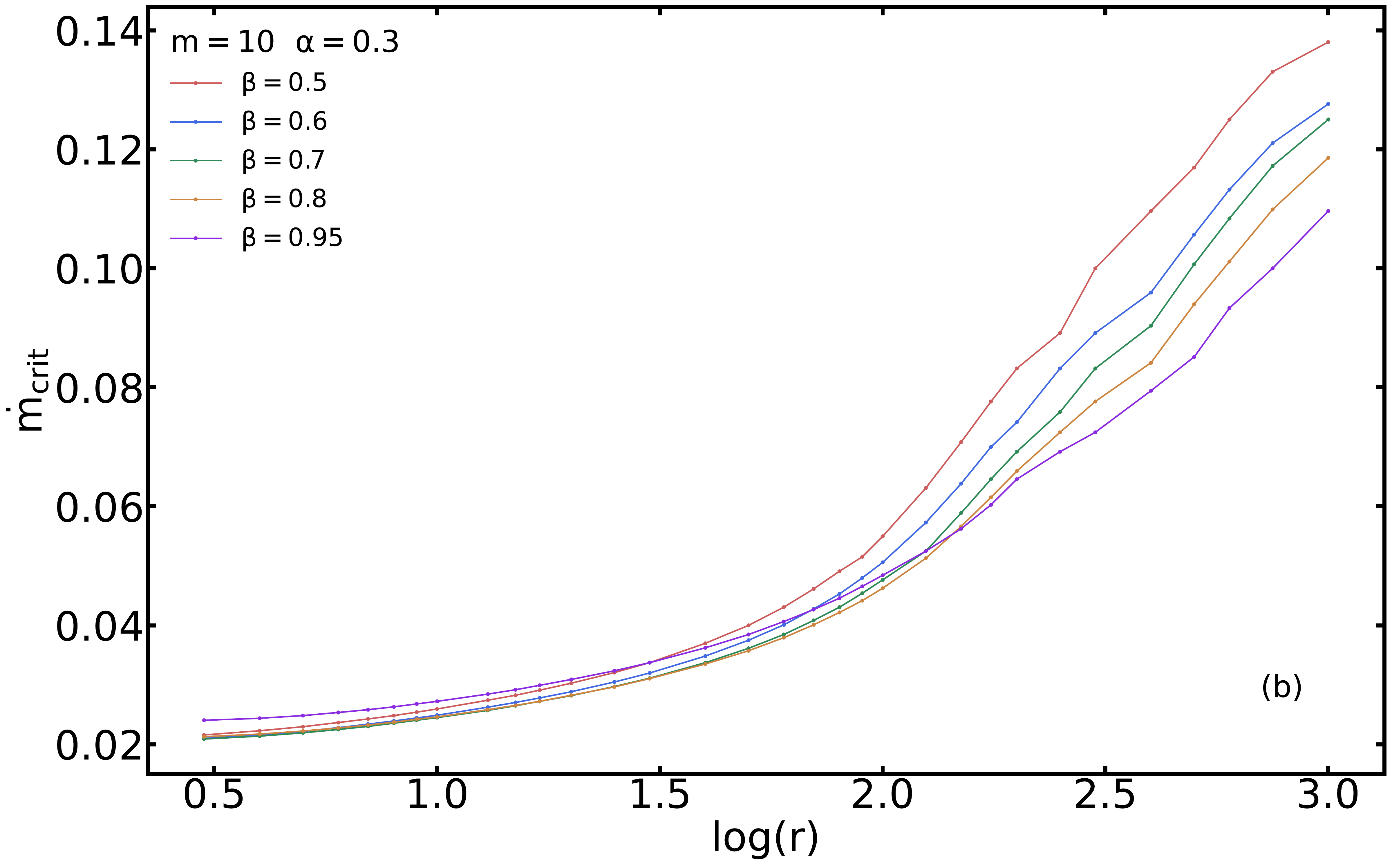}
	\vspace{-0.3cm}
    \caption{Panel (a): The advected fraction of the viscously dissipated energy $f$ as a function of $\dot{m}$ for different $\beta$ for given radii. Panel (b): The critical accretion rate $\dot m_{\rm crit}$ as a function of $r$ for different $\beta$. In the calculations, $m=10$ and $\alpha=0.3$ are adopted.}
\label{theeffectofbeta}
\end{figure*}

In panel (a) of Figure \ref{theeffectofbeta}, for some specific radii, we plot the advected fraction of the viscously dissipated energy $f$ as a function of $\dot{m}$ for different $\beta$.
In the calculations, $m=10$ and $\alpha=0.3$ are adopted. We search for $\dot m_{\rm crit}$ as described in Section \ref{sec31}.
For example, for a radius $r=3$, $\dot m_{\rm crit}=0.022$ for $\beta=0.5$, $\dot m_{\rm crit}=0.021$ for $\beta=0.7$, and $\dot m_{\rm crit}=0.024$ for $\beta=0.95$.  
It is clear that, at $r=3$, the effect of $\beta$ on $\dot m_{\rm crit}$ is very weak, and $\dot m_{\rm crit}$ changes roughly by a factor of $14 \%$ for changing the value of $\beta$ from 0.5 to 0.95. 
The case for $r\lesssim 90$ is similar to that of $r=3$, that there is only very little change of $\dot m_{\rm crit}$ for changing the value of $\beta$ at a fixed radius.
One can refer to Table \ref{tablebeta} for $\dot m_{\rm crit}$ for different $\beta$ as taking $r=10, 30, 90$ respectively for details.
However, in panel (a) of Figure \ref{theeffectofbeta}, we note a very interesting finding that the value of $f$ corresponding to $\dot m_{\rm crit}$ obviously decreases with increasing $\beta$ from 0.5 to 0.95 for a fixed radius. 
Meanwhile, we note that the value of $f$ very weakly depends on $r$. 
The mean values of $f$ for different radii $r=3, 10, 30, 90$ are $0.24, 0.17, 0.04$ for $\beta=0.5, 0.75$ and $0.95$ respectively. 

In panel (b) of Figure \ref{theeffectofbeta}, we plot $\dot m_{\rm crit}$ as a function of $r$ for different $\beta$ (with more radii calculations for $\dot m_{\rm crit}$ added).
In the calculations, $m=10$ and $\alpha=0.3$ are adopted. 
As we can see for the radii of $r\lesssim 90$, the effect of $\beta$ on the relation between $\dot m_{\rm crit}$ and $r$ is weak and nearly can be neglected. 
However, for $r \gtrsim 90$, with increasing $r$, the effect of $\beta$ on the relation between $\dot m_{\rm crit}$ and $r$ becomes obvious, and at $r=1000$, the change of $\dot m_{\rm crit}$ can be a factor of $25\%$ for the change of $\beta$ from 0.5 to 0.95.

\begin{table}
	\caption{The $\dot{m}_{\rm crit}$ for different  $\beta$ at a radius $r$.}
	\label{tablebeta}
	\begin{center}
	\begin{threeparttable}
    \renewcommand\arraystretch{1.2}
	\setlength{\tabcolsep}{15.0pt}
	\begin{tabularx}{0.47\textwidth}{ccccc}
	\hline
    \multicolumn{1}{c}{\multirow{2}{*}{$\beta$}} & \multicolumn{4}{c}{$\dot{m}_{\rm crit}$} \\ 
    & r=3 & r=10 & r=30 & r=90 \\  \hline
    0.5  &  0.022  & 0.026  & 0.034  & 0.052  \\
    0.7  &  0.021  & 0.024  & 0.031  & 0.045  \\
    0.95 &  0.024  & 0.027  & 0.034  & 0.047   \\
     \hline
	\end{tabularx}
	\end{threeparttable}
\end{center}
\end{table}

\subsection{The effect of the viscosity parameter $\alpha$ on $\dot{m}_{\rm{crit}}$} \label{sec34}
\begin{figure*}
	\centering
    \includegraphics[width=0.47\textwidth]{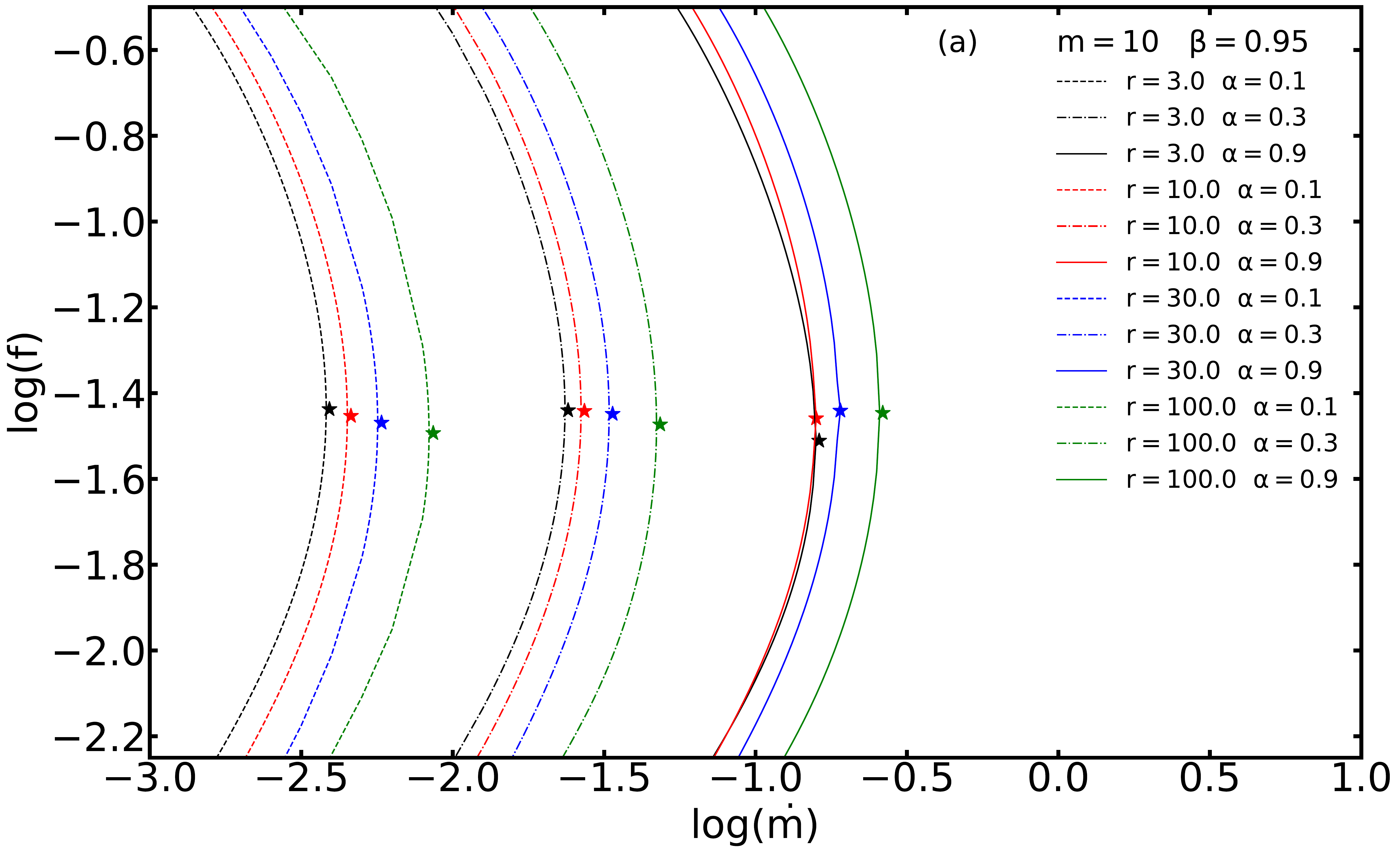}
    \includegraphics[width=0.47\textwidth]{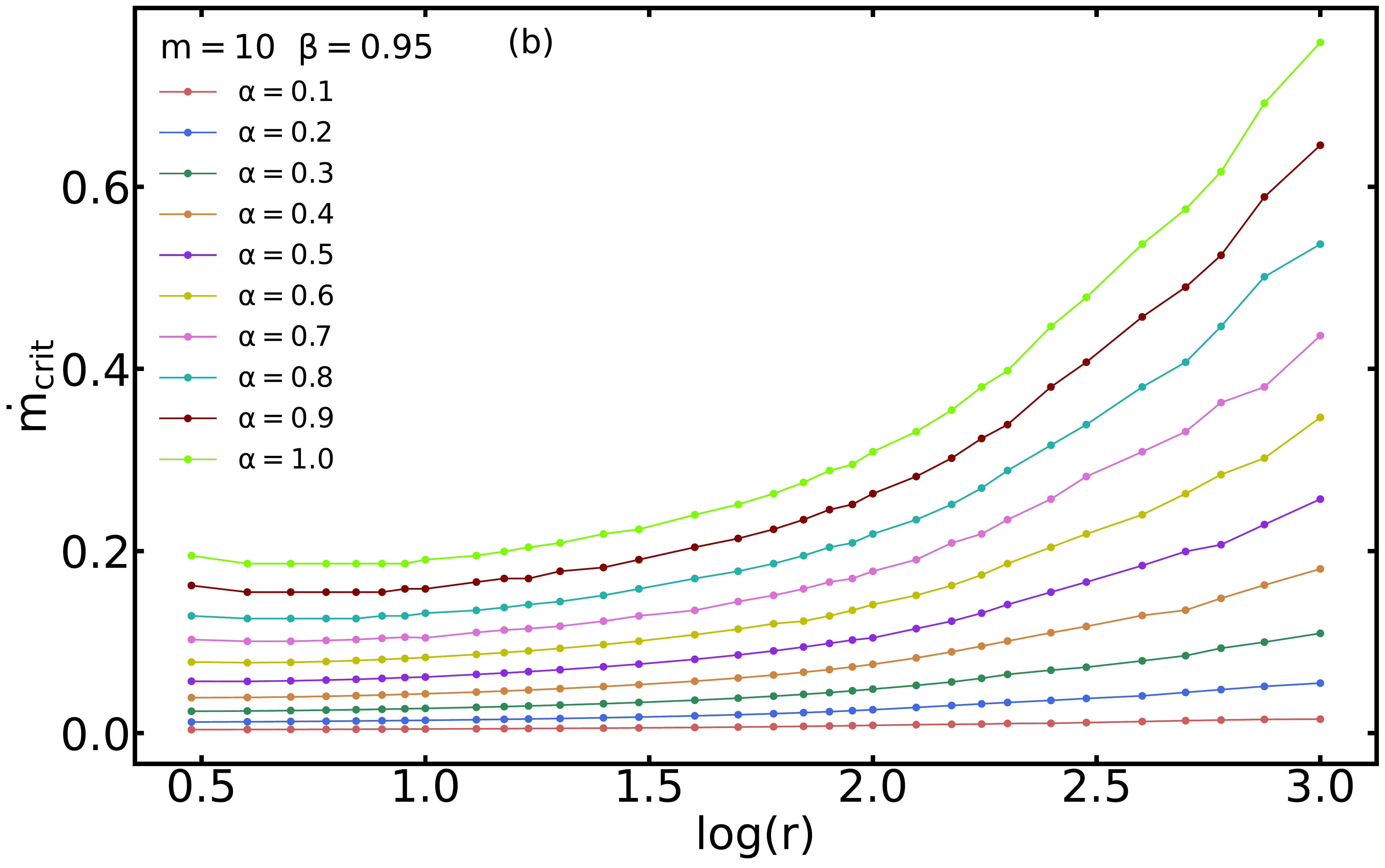}
	\vspace{-0.3cm}
        \caption{Panel (a): The advected fraction of the viscously dissipated energy $f$ as a function of $\dot{m}$ for different $\alpha$ for given radii. Panel (b): The critical accretion rate $\dot m_{\rm crit}$ as a function of $r$ for different $\alpha$. In the calculations, $m=10$ and $\beta=0.95$ are adopted.}
\label{theeffectofalpha}
\end{figure*}

In panel (a) of Figure \ref{theeffectofalpha}, for some specific radii, we plot the advected fraction of the viscously dissipated energy $f$ as a function of $\dot{m}$ for different $\alpha$.
In the calculations, $m=10$ and $\beta=0.95$ are adopted. We also search for $\dot m_{\rm crit}$ as described in Section \ref{sec31}. 
For example, for a radius $r=3$, $\dot m_{\rm crit}=0.004$ for $\alpha=0.1$, $\dot m_{\rm crit}=0.024$ for $\alpha=0.3$, and $\dot m_{\rm crit}=0.162$ for $\alpha=0.9$. 
It is clear that $\dot m_{\rm crit}$ increases with increasing $\alpha$, and there is an increase of $\dot m_{\rm crit}$ by a factor of $\sim 39.5$ for taking $\alpha=0.9$ compared with taking $\alpha=0.1$. 
The cases for other radii are similar to that of $r=3$, that there is an obvious increase of $\dot m_{\rm crit}$ for an increase of $\alpha$.
Specifically, for the radii $r=30, 100, 300, 1000$, there is an increase of $\dot m_{\rm crit}$ by a factor of $\sim 30.8$, $\sim 28.2$, $\sim 32.9$, $\sim 42.1$ for $\alpha$ for taking $\alpha=0.9$ compared with taking $\alpha=0.1$.
One can also refer to Table \ref{tablealpha} for more detailed data of $\dot m_{\rm crit}$ with an interval of $\Delta{\alpha}=0.1$ ranging from  $\alpha=0.1$ to $\alpha=1.0$.

In panel (b) of Figure \ref{theeffectofalpha}, we plot $\dot m_{\rm crit}$ as a function of $r$ for different $\alpha$ (with more radii calculations for $\dot m_{\rm crit}$ added).
In the calculations, $m=10$ and $\beta=0.95$ are adopted. 
As a whole, for a fixed value of $\alpha$, $\dot m_{\rm crit}$ increases with increasing $r$.
Meanwhile, $\dot m_{\rm crit}$ as a function of $r$ systematically shifts upward with increasing the value of $\alpha$.

\begin{table}
	\caption{The $\dot{m}_{\rm crit}$  for different  $\alpha$ at a radius $r$.}
	\label{tablealpha}
	\begin{center}
	\begin{threeparttable}
    \renewcommand\arraystretch{1.2}
	\setlength{\tabcolsep}{10.0pt}
	\begin{tabularx}{0.47\textwidth}{cccccc}
	\hline
    \multicolumn{1}{c}{\multirow{2}{*}{$\alpha$}} & \multicolumn{5}{c}{$\dot{m}_{\rm crit}$} \\ 
    & r=3 & r=30 & r=100 & r=300 & r=1000\\  \hline
    0.1  &  0.004  & 0.006  & 0.009  & 0.012 & 0.015 \\
    0.2  &  0.012  & 0.018  & 0.026  & 0.038 & 0.055 \\
    0.3  &  0.024  & 0.034  & 0.048  & 0.072 & 0.110  \\
    0.4  &  0.039  & 0.053  & 0.076  & 0.117 & 0.180  \\
    0.5  &  0.057  & 0.076  & 0.105  & 0.166 & 0.257  \\
    0.6  &  0.078  & 0.101  & 0.141  & 0.219 & 0.347 \\
    0.7  &  0.103  & 0.129  & 0.178  & 0.282 & 0.437 \\
    0.8  &  0.129  & 0.159  & 0.219  & 0.339 & 0.537 \\
    0.9  &  0.162  & 0.191  & 0.263  & 0.407 & 0.646  \\
    1.0  &  0.195  & 0.224  & 0.309  & 0.479 & 0.759  \\ \hline
	\end{tabularx}
	\end{threeparttable}
\end{center}
\end{table}

\subsection{The dependence of $\dot m_{\rm tr: h\rightarrow s}$ and $\dot m_{\rm tr: s\rightarrow h}$ on $\alpha$} \label{sec35}
As shown in Section \ref{sec32}, the effect of BH mass $m$ on the relation between $\dot m_{\rm crit}$ and $r$ is very weak and nearly can be neglected. 
Meanwhile, we can see from Section \ref{sec33}, for $r\lesssim 90$, the effect of the magnetic parameter $\beta$ on the relation between $\dot m_{\rm crit}$ and $r$ is also very weak and nearly can be neglected.
In Section \ref{sec34}, we show that the effect of the viscosity parameter $\alpha$ on the relation between $\dot m_{\rm crit}$ and $r$ is very obvious. 
In this paper, we only consider the effect of $\alpha$ on $\dot m_{\rm crit}$.

For a given value of $\alpha$, we define the minimum $\dot m_{\rm crit}$ (roughly at ISCO of a BH) as the hard-to-soft transition rate $\dot m_{\rm tr: H\rightarrow S}$, above which BH will gradually transit from the low/hard state (via the intermediate state) to the high/soft state.
We define $\dot m_{\rm crit}$ at $30R_{\rm S}$ 
\footnote{For a non-rotating BH, the fraction of the integrated disk luminosity from $3R_{\rm S}$ to 30$R_{\rm S}$ to the full disk luminosity is $\sim 75\%$. 
In this case, the emission is dominated by the disk emission, and the spectral state is defined as the high/soft state (\citealt{remillard2006}.)} 
as the soft-to-hard transition rate $\dot m_{\rm tr: S\rightarrow H}$, below which BH will gradually transit from the high/soft state (via the intermediate state) to the low/hard state. 

Based on the numerical results shown in panel (b) of Figure \ref{theeffectofalpha}, we get the fitting formulae of $\dot m_{\rm tr: H\rightarrow S}$ and $\dot m_{\rm tr: S\rightarrow H}$ as functions of $\alpha$, which are as follows,
\begin{equation}\label{EQMDOTCRIT_ALPHA}
\begin{aligned}
&\dot m_{\rm tr: H\rightarrow S}=0.183\alpha^{1.677},\\
&\dot m_{\rm tr: S\rightarrow H}=0.226\alpha^{1.585}.
\end{aligned}
\end{equation}

\section{Comparison with Observation}\label{sec4}
\subsection{The dependence of transition luminosities $l_{\rm tr: H\rightarrow S}$ and $l_{\rm tr: S\rightarrow H}$ on $\alpha$} \label{sec41}
In order to more easily compare with observations, we transform the relation between the transition rate of $\dot m_{\rm tr: H\rightarrow S}$ and $\alpha$, as well as the relation between the transition rate $\dot m_{\rm tr: S\rightarrow H}$ and $\alpha$ in equation (\ref{EQMDOTCRIT_ALPHA}) into $l_{\rm tr: H\rightarrow S}$ and $\alpha$, as well as $l_{\rm tr: S\rightarrow H}$ and $\alpha$. 
Here, $l_{\rm tr: H\rightarrow S}$ is defined as the transition luminosity from the low/hard state (via the intermediate state) to the high/soft state in units of $L_{\rm Edd}$, and $l_{\rm tr: S\rightarrow H}$ is defined as the transition luminosity from the high/soft state (via the intermediate state) to the low/hard state in units of $L_{\rm Edd}$.  

More generally, we define the dimensionless luminosity of ADAF $l$ as $l={L\over L_{\rm Edd}}$, which can be further expressed further as,
\begin{equation}\label{equ:l}
\begin{aligned}
l={L\over L_{\rm Edd}}={\eta \dot M c^2\over 0.1\dot M_{\rm Edd}c^2}=\big({\eta \over 0.1}\big)\dot m,
\end{aligned}
\end{equation}
where $\eta$ is the radiative efficiency of ADAF. 
As has been discussed in several previous papers (\citealt{narayan1994, narayan1995b}), in a wide range of $\dot m$ much less than $\dot m_{\rm crit}$, the radiative efficiency $\eta$ of ADAF is less than 0.1, and $\eta$ dramatically decreases with decreasing $\dot m$.
Meanwhile, it was found that $\eta$ increases with increasing $\dot m$.
In particular, when $\dot m$ approaches to $\dot m_{\rm crit}$, the radiative efficiency $\eta$ of ADAF will approach very close to the radiative efficiency of the standard disk $0.1$ for a non-rotating BH (\citealt{xie2012}).

In this paper, we focus on the stage of $\dot m_{\rm tr: H\rightarrow S}$ and $\dot m_{\rm tr: S\rightarrow H}$, at both of which $\eta$ is very close to $0.1$.
So we can easily transform equation (\ref{EQMDOTCRIT_ALPHA}) as follows,
\begin{equation}\label{EQLTR_ALPHA}
\begin{aligned}
&l_{\rm{tr:H\rightarrow S}}=0.183\alpha^{1.677},\\
&l_{\rm{tr:S\rightarrow H}}=0.226\alpha^{1.585}.
\end{aligned}
\end{equation} 

\subsection{Sources selection and luminosities correction}\label{sec42}
\subsubsection{Sources selection}\label{sec421}
We search for sources with measured transition luminosity between the low/hard state and the high/soft state from the literature. 
One can refer to Table \ref{tabledata} for more details. 
The data for the hard-to-soft transition are from \cite{yu2004,yu2007,yu2009,bhuvana2021}, and the data for the soft-to-hard state transition are from \cite{maccarone2003a,motlagh2019,bhuvana2021}.
In total, we have $26$ observations from $10$ sources. 
All these sources have dynamically measured BH mass, and relatively accurate distance measures, both of which are important for measuring the dimensionless transition luminosities of $l_{\rm tr}$ defined as $l_{\rm tr}={L_{\rm tr}\over L_{\rm Edd}}$.  
The definition of $L_{\rm tr}$ depends on the specific energy band. 
One can refer to Section \ref{sec422} and Section \ref{sec423} for the luminosity correction for details.
Here, we should note that several sources are not included in our sample, since they are either without BH mass measures or distance measures.
For example, H$1743-322$ is lack of BH mass measure (\citealt{motlagh2019}), XTE J$1650-500$, XTE J$1720-318$, XTE J$1748-288$, XTE J$1817-330$, XTE J$1908+094$ and XTE J$1752-223$ (\citealt{motlagh2019}), as well as XTE J$1856+053$ and Cyg X-3 from (\citealt{yu2009}) are either lack of BH mass measures or distance measures or are even lack of both BH mass and distance measures.

\subsubsection{Lumonosities correction in Maccarone (2003) and Vahdat Motlagh et al. (2019)} \label{sec422}

As we mentioned in Section \ref{sec421}, the definition of the transition luminosities $L_{\rm tr}$ depends on the specific energy band.
In \cite{maccarone2003a}, the authors corrected the observed X-ray luminosities to a wider energy range of $0.5$ keV $- 10$ MeV by assuming the form of the spectrum as $dN/dE \sim E^{-1.8}\rm{exp}^{-E/200\rm{keV}}$.
The energy band of $0.5$ keV $- 10$ MeV safely covers the emission of both the soft X-ray component and the hard X-ray component that may affect the transition luminosities calculations.
\cite{motlagh2019} corrected the observed X-ray luminosities to an energy range of $0.5 - 200$ keV as the transition luminosities. 
Meanwhile, the same spectral form of $dN/dE \sim E^{-1.8}\rm{exp}^{-E/200\rm{keV}}$ was also used to re-calculate the X-ray luminosity in $0.5$ keV $- 10$ MeV, which agreed with the data corrected above in $10$ percent.
In this paper, we directly use the corrected data of \cite{maccarone2003a} and \cite{motlagh2019} as the transition luminosities, which are all listed in Table \ref{tabledata} for clarity.

\subsubsection{Luminosities correction for Yu et al. (2004, 2007); Yu \& Yan (2009); Bhuvana et al. (2021)}\label{sec423}
In \cite{yu2004,yu2007,yu2009}, the transition luminosities are calculated in \emph{Swift}/BAT band of $15-50$ keV, and in \cite{bhuvana2021}, the transition luminosities are calculated in \emph {NICER} band of $0.3-10$ keV.
We simply extrapolate the observed X-ray luminosities in the \emph {Swift}/BAT band of $15-50$ keV and the \emph {NICER} band of $0.3-10$ keV to the energy band of $0.1-200$ keV by assuming the spectral form of $dN/dE \sim E^{-1.8}\rm{exp}^{-E/200\rm{keV}}$ as the transition luminosities. 
One can refer to Table \ref{tabledata} for our corrected transition luminosities based on the data of \cite{yu2004,yu2007,yu2009,bhuvana2021} for details.

\subsection{Constraining $\alpha$ with the observed transition luminosities}\label{sec43}
\begin{figure}
	\centering
     \includegraphics[width=0.47\textwidth]{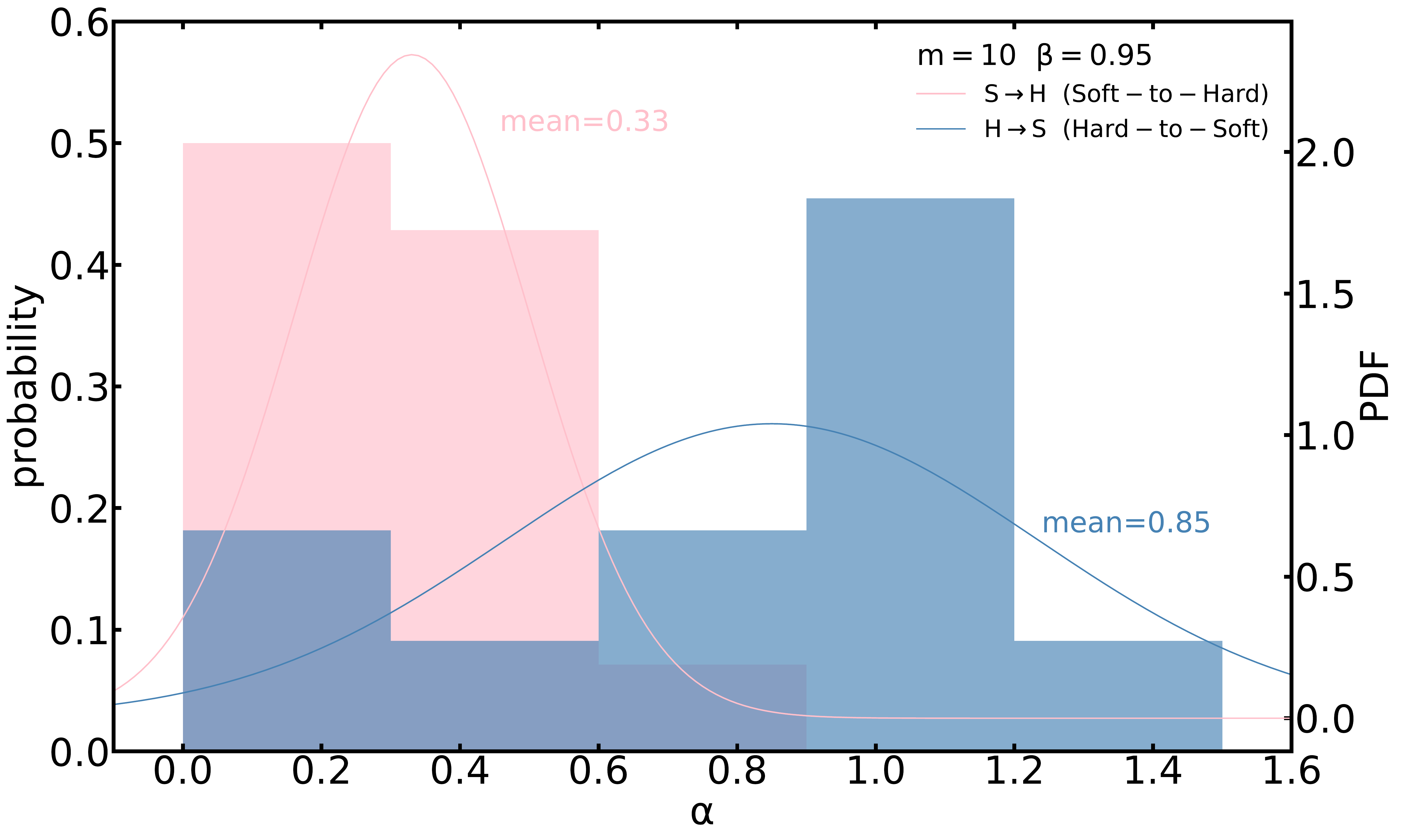}
	\vspace{-0.3cm}
    \caption{The distribution of the viscosity parameter $\alpha$. The blue histogram is the distribution of $\alpha$ for the observations of the hard-to-soft transition,
    and the pink histogram is the distribution of $\alpha$ for the observations of the soft-to-hard transitions. The mean value of $\alpha$ for the hard-to-soft transition is 0.85,
    and the mean value of $\alpha$ for the soft-to-hard transition is 0.33. The blue curve refers to the data of the hard-to-soft transition smoothed with normal distribution, and the pink curve refers to the data of the soft-to-hard transition smoothed with normal distribution. } 
    \label{mdottrandalpha}
\end{figure}

With the observed transition luminosities of both the hard-to-soft transition and the soft-to-hard transition, we solve the first equation in Equation (\ref{EQLTR_ALPHA}) for the value of $\alpha$ of the hard-to-soft transition and solve the second equation in Equation (\ref{EQLTR_ALPHA}) for the value of $\alpha$ of the soft-to-hard transition separately. 
All the derived values of $\alpha$ are listed in the last column of Table \ref{tabledata}. 
In Figure \ref{mdottrandalpha}, we plot the distribution of $\alpha$ for the observations of the hard-to-soft transition and the distribution of $\alpha$ for the observations of the soft-to-hard transition, respectively. 
It is found that the mean value of $\alpha$ of the hard-to-soft transition is $\alpha \sim 0.85$, and the mean value of $\alpha$ of the soft-to-hard transition is $\alpha \sim 0.33$.
It is clear that the data suggest a larger mean value of $\alpha$ for the hard-to-soft transitions and a smaller value of $\alpha$ for the soft-to-hard transitions to match the transition luminosities.

In order to test the discrepancy of the value of $\alpha$ between the hard-to-soft transition and the soft-to-hard transition, we calculate the mean value of $\alpha$ for the hard-to-soft transition and the soft-to-hard transition respectively for a single source. 
One can refer to Table \ref{tablealphamean} for details.
For example, for GX $339-4$, one of the most studied BH low-mass X-ray binaries, the mean value of $\alpha$ for the hard-to-soft transition is $0.95$, and the mean value of $\alpha$ for the soft-to-hard transitions is $0.45$. 
For other sources, such as GRO J$1655-40$, XTE J$1550-564$, LMC X-3, the mean value of $\alpha$ for the hard-to-soft transition is also larger than that of the soft-to-hard transition respectively.
The only exception is Cyg X-1, for which the value of $\alpha$ for the hard-to-soft transition is slightly smaller than that of the soft-to-hard transitions. 
The physical reasons for such a discrepancy of the mean value of $\alpha$ for the hard-to-soft transition and the soft-to-hard transition between Cyg X$-1$ and other sources are not clear, but it is very possible that the accretion mode (wind accretion) in Cyg X$-1$ is different from that of other sources (Roche lobe accretion) (see the discussion of \citealt{taam2018}).  
  
\begin{table*}
	\caption{The sources with state transition identified, parameters used, the state transition luminosities and the viscosity parameter calculated.}
	\label{tabledata}
	\begin{center}
	\begin{threeparttable}
    \renewcommand\arraystretch{1.3}
	\setlength{\tabcolsep}{5.5pt}
	\begin{tabularx}{\textwidth}{lcccclcr}
		\hline
		{Source} & {Mass} & {Distance} & {Date} & {State } & 	\multicolumn{1}{c}{Transition Luminosity}  & \multicolumn{1}{c}{Reference} &\multicolumn{1}{c}{Viscosity}  \\ 
		&($M_{\bigodot}$)&(kpc)&(MJD)& Transition & \multicolumn{1}{c}{$(l_{\rm{tr}}=L_{\rm{tr}}/L_{\rm{Edd}})$\tnote{a}}&&{Parameter$(\alpha)$}\\
		\hline
		{GX 339-4} & {$5.8\pm0.5$}   & {$5.6$} & {1998(50711)}  &{$\rm H\rightarrow S$}\tnote{b} & {$0.0656^{+0.0264}_{-0.0108}$}\tnote{ c} &\cite{yu2007} &{$0.54^{+0.13}_{-0.05}$} \\
		&{$5.8\pm0.5$}  &  {$5.6$} & {2003(52394)} &{$\rm H\rightarrow S$} & {$0.3505^{+0.1410}_{-0.0579}$}\tnote{ c} &\cite{yu2007}&{$1.47^{+0.35}_{-0.15}$} \\
		&{$9.0\pm1.4$}  & {$8.4\pm0.9$} & {2003(52717)}  &{$\rm S\rightarrow H$}\tnote{b} & {$0.0539\pm0.0147$}\tnote{ d} &\cite{motlagh2019}&{$0.40\pm0.07$}\\ 
		&{$5.8\pm0.5$}  &  {$5.6$} & {2005(53225)} &{$\rm H\rightarrow S$} & {$0.1043^{+0.0420}_{-0.0172}$}\tnote{ c} &\cite{yu2007}&{$0.72^{+0.17}_{-0.07}$}\\ 
		&{$9.0\pm1.4$}  &  {$8.4\pm0.9$}  & 2005(53457) &{$\rm S\rightarrow H$} & {$0.0752\pm0.0198$}\tnote{ d}  &\cite{motlagh2019}&{$0.50\pm0.08$}\\ 
		& {$5.8\pm0.5$} &  {$5.6$} & {2007(54140)} &{$\rm H\rightarrow S$} & {$0.2053^{+0.0826}_{-0.0339}$}\tnote{ c} & \cite{yu2009}&{$1.07^{+0.26}_{-0.11}$}\\ 
		& {$9.0\pm1.4$} &  {$8.4\pm0.9$} & {2007(54234)} &{$\rm S\rightarrow H$} &{$0.0678\pm0.0307$}\tnote{ d}& \cite{motlagh2019}&  {$0.47\pm0.13$} \\
		&{$9.0\pm1.4$}  &  {$8.4\pm0.9$} & {2011(55594)} &{$\rm S\rightarrow H$} & {$0.0588\pm0.020$}\tnote{ d} & \cite{motlagh2019}&{$0.43\pm0.09$}\\
		\hline
		{GRO J1655-40} &  {$6.3\pm0.5$}  & {$3.3\pm0.2$}   & {1997(50674)}  & {$\rm S\rightarrow H$}&  {$0.0095\pm0.0024$}\tnote{ e} & \cite{maccarone2003a}&{$0.14\pm0.02$}\\
		& {$6.3\pm0.5$}  & {$3.2$}   & {2005(53437)}  & {$\rm H\rightarrow S$}&  {$0.0132^{+0.0053}_{-0.0022}$}\tnote{ c} & \cite{yu2009}&{$0.21^{+0.05}_{-0.02}$}\\  
	    & {$5.4\pm0.3$}  &  {$3.2\pm0.2$}   & {2005(53627)} & {$\rm S\rightarrow H$}  & {$0.0251\pm0.0030$}\tnote{ d} &\cite{motlagh2019}&{$0.25\pm0.02$}\\
		\hline
		{XTE J1550-564} & {$10.8$} & {-}  & {1999(51063)}  &{$\rm H\rightarrow S$}  &  {$0.2183^{+0.0878}_{-0.0361}$}\tnote{ c} & \cite{yu2004}&{$1.11^{+0.15}_{-0.06}$} \\
	    & {$9.1\pm0.6$}  & {$4.4\pm0.5$}  & {1999(51304)} & {$\rm S\rightarrow H$} & {$0.0031\pm0.0008$}\tnote{ d} & \cite{motlagh2019}&{$0.07\pm0.01$}\\ 
	    & {$10.8$} & {-} & {2000(51654)} &{$\rm H\rightarrow S$} &  {$0.1262^{+0.0508}_{-0.0209}$}\tnote{ c}  & \cite{yu2004}&{$0.80^{+0.19}_{-0.08}$} \\ 
		& {$9.1\pm0.6$}  & {$4.4\pm0.5$} & {2000(51675)} &{$\rm S\rightarrow H$} & {$0.0530\pm0.0109$}\tnote{ d} & \cite{motlagh2019}&{$0.40\pm0.05$}\\ 
		\hline
		{4U 1543-47}  & {$9.4\pm2.0$}  & {$7.5\pm0.5$} & {2002(52473)} & {$\rm S\rightarrow H$}& {$0.1265\pm0.0206$}\tnote{ d} & \cite{motlagh2019}&{$0.69\pm0.07$}\\ 
		\hline
		{XTE J1650-500}  & {$8\pm1.5$}  & {$8\pm2$}  & {2001(52231)} &{$\rm S\rightarrow H$}  &  {$0.1329\pm0.0664$}\tnote{ d} & \cite{motlagh2019}&{$0.72\pm0.23$}\\ 
		\hline
		{GRS 1915-105}  & {$14\pm4$} & { $11.2-12.5$}  & {2005(53460)} &{$\rm H\rightarrow S$} &  {$0.2086^{+0.0839}_{-0.0345}$}\tnote{ c} & \cite{yu2009}&{$1.08^{+0.26}_{-0.11}$}\\ 
		& {$14\pm4$} & {$11.2-12.5$}  & {2007(54267)}   &{$\rm H\rightarrow S$} &  {$0.2351^{+0.0946}_{-0.0389}$}\tnote{ c} & \cite{yu2009}&{$1.16^{+0.28}_{-0.11}$} \\ 
		\hline
		{Nova Muscae 1991}  & {$7.0\pm0.6$} & { $5.1\pm0.7$}  & {1991(48399)} &{$\rm S\rightarrow H$} &  {$0.031\pm0.0109$}\tnote{ e} & \cite{maccarone2003a}&{$0.29\pm0.06$}\\
		\hline
		{GS 2000+251}  & {$8.5\pm1.5$} & { $2.0\pm1.0$}  & {1988(47409)} &{$\rm S\rightarrow H$} &  {$0.0069\pm0.0069$}\tnote{ e} & \cite{maccarone2003a}&{$0.11\pm0.07$}\\
		\hline
		{Cyg X-1} & {$10.1$} & {$2.1$} & {1996(50200)} & {$\rm H\rightarrow S$} &  {$0.0126^{+0.0050}_{-0.0021}$} \tnote{ d} & \cite{yu2009}&{$0.20^{+0.02}_{-0.05} $}\\
        & {$13.0\pm3.0$} & {$2.5\pm0.5$} & {1996(50328)} & {$\rm S\rightarrow H$} &  {$0.0280\pm0.0140$}\tnote{ d} & \cite{maccarone2003a}&{$0.27\pm0.08$}\\		
		\hline
		{LMC X-3}  & {$9.5\pm2.0$} & {$51\pm1.0$}  & {1998(50962)} &{$\rm S\rightarrow H$} &  {$0.0140\pm0.0041$}\tnote{ e} & \cite{maccarone2003a}&{$0.17\pm0.03$}\\
		& {$6.48$} & {$48.1$}  & {2018(58463)} &{$\rm H\rightarrow S$} &  {$0.1772^{+0.0311}_{-0.0195}$}\tnote{ c} & \cite{bhuvana2021}&{$0.98^{+0.10}_{-0.06}$}\\
		& {$6.48$} & {$48.1$}  & {2018(58532)} &{$\rm S\rightarrow H$} &  {$0.0630^{+0.0110}_{-0.0069}$}\tnote{ c} & \cite{bhuvana2021}&{$0.45^{+0.05}_{-0.03}$}\\	
		\hline
	\end{tabularx}
		\begin{tablenotes}
		\footnotesize 
		\item[a.] $l_{\rm{tr}}$ indicates the transition luminosity in units of $L_{\rm Edd}$, and in this column including both the hard-to-soft transition luminosity $l_{\rm tr: H\rightarrow S}$ and the soft-to-hard transition luminosity $l_{\rm tr: S\rightarrow H}$.
		\item[b.] $\rm H\rightarrow S$ indicates the hard-to-soft transition; $\rm S\rightarrow H$ indicates the soft-to-hard transition.
		\item[c.] The corrected transition luminosities based on the data of \cite{yu2004,yu2007,yu2009,bhuvana2021} as described in Section \ref{sec423}
        \item[d.] The transition luminosities from \cite{maccarone2003a}, which are summarized at Section \ref{sec422}.
	    \item[e.] The transition luminosities from \cite{motlagh2019}, which are summarized at Section \ref{sec422}. 
		\end{tablenotes}
	\end{threeparttable}
\end{center}
\end{table*}

\begin{table}
    \caption{The average $\overline \alpha$ for the hard-to-soft transition and the soft-to-hard transition for different sources.}
    \label{tablealphamean}
    \begin{center}
    \begin{threeparttable}
    \renewcommand\arraystretch{1.3}
    \setlength{\tabcolsep}{10.0pt}
\begin{tabular}{lll}
    \hline
    Source  & $\overline \alpha_{H \rightarrow S}$ & $\overline{\alpha}_{S \rightarrow H}$ \\ 
    \hline
    GX 339-4         & 0.95 & 0.45  \\
    GRO J1655-40     & 0.21 & 0.20  \\
    XTE J1550-564    & 0.96 & 0.24  \\
    4U 1543-47       &  -   & 0.69   \\
    XTE J1650-500    &  -   & 0.72  \\
    GRS 1915-105     & 1.12 & - \\
    Nova Muscae 1991 &  -   & 0.29   \\
    GS 2000+251      &  -   & 0.11    \\
    Cyg X-1          & 0.20 & 0.27  \\
    LMC X-3          & 0.98 & 0.31    \\ 
    \hline
\end{tabular}	
\end{threeparttable}
\end{center}
\end{table}

\section{Discussion}\label{sec5}
\subsection{Geometry of the accretion flow}\label{geometry}
One of the most important findings in this paper is that $\dot m_{\rm crit}$ increases with increasing $r$, based on which we have the geometry of the accretion flow as shown in Figure \ref{geometry}. 
As we can see in the left panel of Figure \ref{geometry}, when the accretion rate is greater than the minimal critical accretion rate (roughly at the ISCO of a non-rotating BH), part of the ADAF gas will collapse to form a weak disk in the inner region. 
The source starts to transit from the low/hard state (via the intermediate state) to the high/soft state. 
When the accretion rate is greater than the critical accretion rate at $30R_{\rm S}$, we define the source to enter the high/soft state as the emission of accretion flow is dominated by the accretion disk. 
In the opposite direction, as can be seen in the right panel of Figure \ref{geometry}, if the mass accretion rate decreases less than the critical accretion rate at $30 R_{\rm S}$, the source will gradually transit from the high/soft state (via the intermediate state) to the low/hard state. 
During this process, the size of the inner disk decreases with decreasing mass accretion rate, and finally, when the accretion rate is less than the minimal critical accretion rate, the disk is completely evaporated and the source enters the low/hard state. 
Such a picture described above is very similar to our proposed condensation model of the hot gas, which has been successfully used to explain the geometry and the spectral feature of both BH X-ray binaries (\citealt{meyer2007,liu2007,liu2011,qiao2013,taam2018}) and active galactic nuclei(\citealt{liu2015,liu2017,qiao2017,qiao2018,liu2022} for review).

It is shown from the above statement that the accretion flow is comprised of hot and cold flows, dominated by the ADAF at low accretion rates, while dominated by the thin disk at high accretion rates. 
The hot and cold gases interact with each other during the accretion.
The relative strength of the thin disk and the ADAF is changed by the variation of the gas supply rate, which leads to a change in the spectrum and state transition.
Here we should mention that the hysteresis effect, i.e. the transition luminosity from the low/hard state to the high/soft state is often roughly a few times higher than that from the high/soft state to the low/hard state. 
If such the hysteresis effect can be explained in the framework of ADAF solution, a bigger value of $\alpha$ is needed to match the transition luminosity of the hard-to-soft transition and a smaller value of $\alpha$ is needed to match the transition luminosity of the soft-to-hard transition, which will be discussed in the next Section \ref{valueofalpha}.

Also, we would like to mention that the geometry of the accretion flow and the corresponding physics we proposed in this paper is only derived from the relation between $\dot m_{\rm crit}$ and $r$ predicted by the self-similar solution of ADAF, which was further applied to BH-XRBs for studying the phenomena of spectral state transitions. 
Actually, the initial mass supply from the companion star is not so simple, e.g., which could be in the form of hot wind as described with ADAF in this paper or could be in the form of a cool disk via the Roche lobe overflow.
As for the case of the Roche lobe accretion, the initial disk could be transitioned to a hot gas (such as ADAF) via evaporation or some mechanisms that are not very clear so far. 
Meanwhile, we have to mention that in the case of the Roche lobe accretion, it is unavoidable that matter will be accumulated in the outer region of the accretion disk due to the existence of hydrogen ionization instability (viscous instability), which finally can result in a dramatic increase of the mass accretion rate. 
The effect of the dramatic increase of the mass accretion rate can result in an outburst, during which many physical processes occurred such as the observed spectral state transitions and other complicated spectral evolution and variabilities. 
The research in this paper actually studied the geometry of the accretion flow and the related spectral state transition in BH-XRBs with changing the mass accretion rate in the framework of a self-similar solution of ADAF.

\begin{figure}
	\centering
     \includegraphics[width=0.47\textwidth]{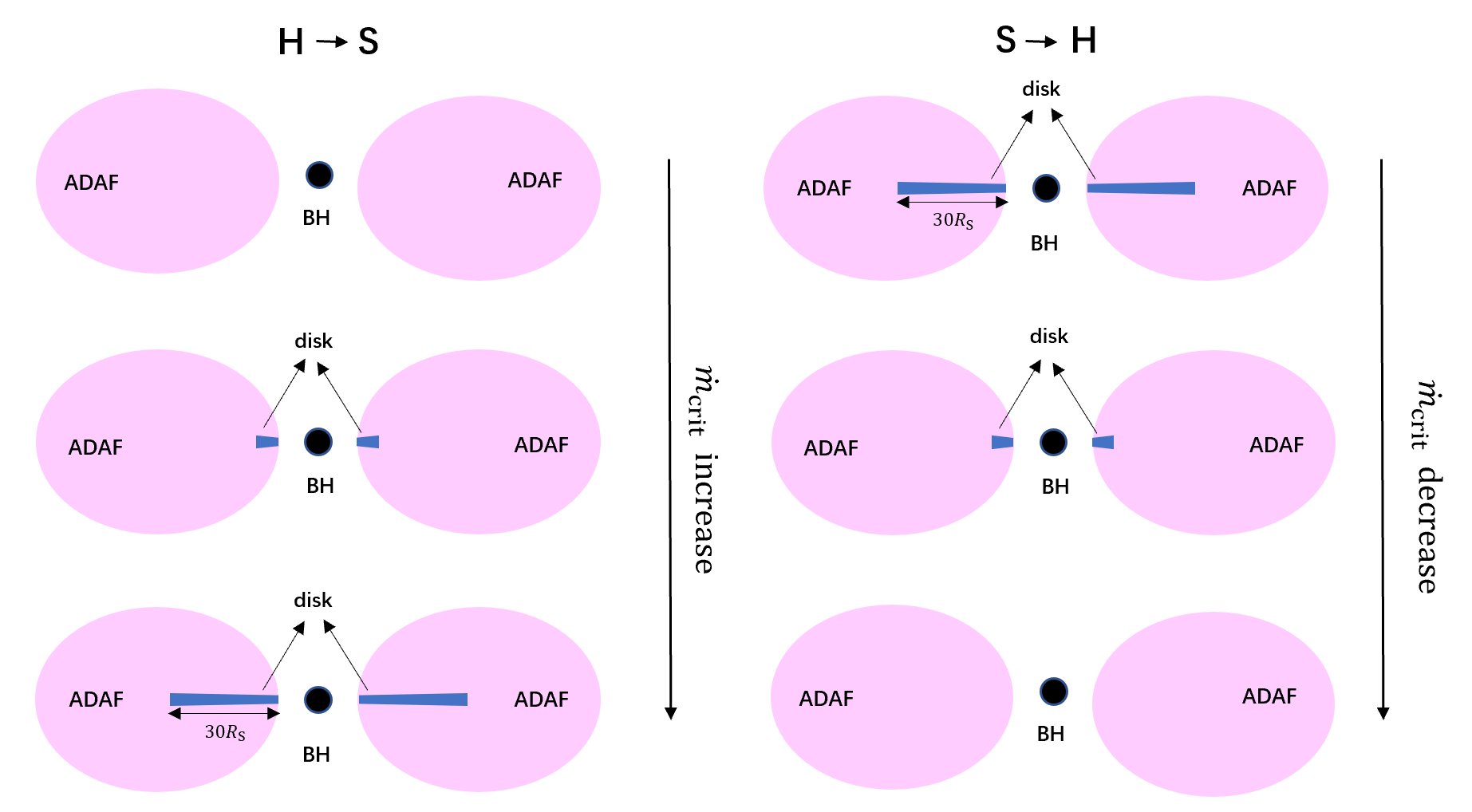}
	\vspace{-0.3cm}
    \caption{Variation of the geometry of accretion flow with the accretion rate. } 
    \label{geometry}
\end{figure}

\subsection{On the value of $\alpha$}\label{valueofalpha}
The mechanism of angular momentum transport is one of the most important processes in the field of accretion theory. 
In \cite{shakura1973}, the authors studied the accretion process in the low-mass X-ray binaries, in which a very important parameter, i.e., the viscosity parameter $\alpha$, is introduced for describing the complicated mechanism for angular momentum transport.
The $\alpha$ description for the transport of angular momentum has been widely used in the subsequent studies of the accretion flow, such as the famous solutions for lower mass accretion rate, i.e., ADAF and its different variants (\citealt{narayan1994,narayan1995b}), and for higher mass accretion rate, i.e., the slim disk and the super-Eddington accretion (\citealt{katz1977,begelman1979,begelman1982,abramowicz1988,chen1993,ohsuga2005}). 

A great deal of effort has been made to determine the value of $\alpha$ over the last decades. 
In \cite{king2007}, the author summarized both the observational and theoretical estimates of the value of $\alpha$, suggesting a typical range $\alpha \sim 0.1-0.4$ observationally for a fully ionized thin disk(\citealt{pringle1981,hartmann1998,smak1999,dubus2001}), and theoretically a much smaller value of $\alpha$ (roughly one order of magnitude smaller) from numerical simulations (\citealt{gammie2001,rice2005,davis2010,simon2012,bai2013,salvesen2016,hameury2020}).
\cite{kempski2019} obtained $\alpha \sim  0.1$ for ADAF by presenting a systematic shearing-box investigation of the magnetorotational instability (MRI)-driven turbulence in a weakly collisional plasma. 
However, some recent simulations have tended to give larger values (\citealt{kempski2019,hameury2020}). 
As discussed in \cite{king2007}, we should note that both the observed value of $\alpha$ and the value of $\alpha$ of the numerical simulation are uncertain.
In particular, we would like to mention that the value of $\alpha$ obtained from numerical simulations depends on how to do the numerical simulation (e.g. considering the magnitude and configuration of the magnetic field, boundary conditions, etc.), which can make significant changes of the value of $\alpha$.

Several previous theoretical studies have shown that the transition luminosities are very sensitive to the value of $\alpha$, such as the study in the framework of the disk evaporation model (\citealt{meyer2000a,meyer2000b,liu2002,qiao2009,taam2012}), and in the framework of ADAF solutions (\citealt{narayan1994,narayan1995b,mahadevan1997}).
In this paper, we search for the transition luminosities between the low/hard state and the high/soft state (and vice versa) in BH-XRBs from the literature. 
We re-investigate the transition luminosity by searching for $\dot m_{\rm crit}$ for different radii one by one ranging from ISCO of a non-rotating BH to 1000$R_{\rm S}$, defining two quantities, i.e., $\dot m_{\rm tr: H\rightarrow S}$ and $\dot m_{\rm tr: S\rightarrow H}$ corresponding to the mass accretion rate of the transition luminosities from the hard-to-soft transition $l_{\rm tr: H\rightarrow S}$ and the soft-to-hard transition $l_{\rm tr: S\rightarrow H}$ respectively.
Further, fitting formulae of $l_{\rm tr: H\rightarrow S}$ and $l_{\rm tr: S\rightarrow H}$ as a function of $\alpha$ is derived, with which the observed transition luminosity of both the hard-to-soft transition and the soft-to-hard transition we precisely constrained the value of $\alpha$. 
Here, we would like to mention that compared with the very rough estimates for $\dot m_{\rm crit}$ and the corresponding transition luminosity $l_{\rm tr}$ in the previous studies of ADAF solution, two new formulae of $l_{\rm tr: H\rightarrow S}$ and $l_{\rm tr: S\rightarrow H}$ as functions of $\alpha$ are more accurate, which is one of the major points in the present paper.
The constrained value of $\alpha$ for the soft-to-hard transition is $\sim 0.33$, which is roughly in the suggested range of \cite{king2007}. 
The constrained value of $\alpha$ for the hard-to-soft transition is $\sim 0.85$,  which is larger than that of the suggested value in \cite{king2007}. But it is consistent with the suggestions for applying ADAF solution in the luminous hard state of BH-XRBs, e.g. \cite{narayan1996}. We think the constrained larger value of $\alpha \sim 0.85$ during the hard-to-soft transition is reasonable since BHs indeed transit to a more luminous state at the stage of hard-to-soft transition.
Meanwhile, \cite{tetarenko2018} determined the  corresponding viscosity parameters to be $\alpha \sim 0.2-1$ by reporting the results of an analysis of archival X-ray light-curves of twenty-one BH-XRBs outbursts. The viscosity parameters are derived in both outbursts where the source cycles through all the accretion states and those where the source remains only in the hard state. 

\section{Conclusion}\label{sec6}
In this paper, we search for the critical mass accretion rate $\dot m_{\rm crit}$ of the self-similar solution of ADAF for different radii $r$ ranging from ISCO ($3 R_{\rm S}$) of a non-rotating BH outward up to $1000 R_{\rm S}$. 
It is found that, generally, $\dot m_{\rm crit}$ decreases with decreasing $r$. 
We test the effect of BH mass $m$, the magnetic parameter $\beta$ and the viscosity parameter $\alpha$ on the relation between $\dot m_{\rm crit}$ and $r$, and have several new results, which are summarized as follows,

1. The effect of BH mass $m$ on the relation between $\dot m_{\rm crit}$ and $r$ is very weak, and can be neglected.

2. For $r\lesssim 90$, the effect of $\beta$ on the relation between $\dot m_{\rm crit}$ and $r$ is very weak, and can be neglected.  

3. The effect of $\alpha$ on the relation between $\dot m_{\rm crit}$ and $r$ is strong. In this case, we define the minimum $\dot m_{\rm crit}$ (roughly at ISCO) as the hard-to-soft transition rate $\dot m_{\rm tr: H\rightarrow S}$, and  $\dot m_{\rm crit}$ at $30R_{\rm S}$ as the soft-to-hard transition rate $\dot m_{\rm tr: S\rightarrow H}$. 
Meanwhile, we assume that at $\dot m_{\rm tr: H\rightarrow S}$ and $\dot m_{\rm tr: S\rightarrow H}$ the radiative efficiency $\eta$ of ADAF is 0.1.
Then we derived fitting formulae for the transition luminosities of hard-to-soft and soft-to-hard as functions of $\alpha$ respectively, which are listed as follows, 
\begin{equation*}\label{EQLTR_ALPHA}
\begin{aligned}
&l_{\rm{tr:H\rightarrow S}}=0.183\alpha^{1.677},\\
&l_{\rm{tr:S\rightarrow H}}=0.226\alpha^{1.585}.
\end{aligned}
\end{equation*} 
  
Based on the observed transition luminosities of a sample of ten BH-XRBs with 26 observations, including both the hard-to-soft transition and the soft-to-hard transition, we constrain the values of $\alpha$ by solving the above formulae of $l_{\rm{tr: H \rightarrow S}}$ and $l_{\rm{tr: S \rightarrow H}}$ as a function of $\alpha$ respectively. 
It is found that the mean value of $\alpha$ for the hard-to-soft transition is 0.85, and the mean value of $\alpha$ for the soft-to-hard transition is 0.33, which indicates that two classes of $\alpha$ are needed for explaining the observed hysteresis effect of the transition luminosity. 
Finally, we discussed that such a finding of the discrepancy of $\alpha$ between the hard-to-soft transition and the soft-to-hard transition may provide some clues for exploring the accretion physics that occurred in the phase of spectral state transition of accreting stellar-mass BHs.

\section*{Acknowledgements}
JQL thanks Wenfei Yu and Zhen Yan for providing some data and very useful discussions.  JQL thanks Huaqing Cheng and Haonan Yang for helpful discussions. 
This work is supported by the National Natural Science Foundation of China (grants 12173048), NAOC Nebula Talents Program, 
the gravitational wave pilot B (Grant No. XDB23040100), and the Strategic Pioneer Program on Space Science, Chinese Academy of Sciences (Grant No. XDA15052100).

\section*{Data Availability}
The data underlying this article will be shared on reasonable request to the corresponding author.

\bibliographystyle{mnras}
\bibliography{reference}

\begin{thebibliography}{}
\makeatletter
\relax
\def\mn@urlcharsother{\let\do\@makeother \do\$\do\&\do\#\do\^\do\_\do\%\do\~}
\def\mn@doi{\begingroup\mn@urlcharsother \@ifnextchar [ {\mn@doi@}
  {\mn@doi@[]}}
\def\mn@doi@[#1]#2{\def\@tempa{#1}\ifx\@tempa\@empty \href
  {http://dx.doi.org/#2} {doi:#2}\else \href {http://dx.doi.org/#2} {#1}\fi
  \endgroup}
\def\mn@eprint#1#2{\mn@eprint@#1:#2::\@nil}
\def\mn@eprint@arXiv#1{\href {http://arxiv.org/abs/#1} {{\tt arXiv:#1}}}
\def\mn@eprint@dblp#1{\href {http://dblp.uni-trier.de/rec/bibtex/#1.xml}
  {dblp:#1}}
\def\mn@eprint@#1:#2:#3:#4\@nil{\def\@tempa {#1}\def\@tempb {#2}\def\@tempc
  {#3}\ifx \@tempc \@empty \let \@tempc \@tempb \let \@tempb \@tempa \fi \ifx
  \@tempb \@empty \def\@tempb {arXiv}\fi \@ifundefined
  {mn@eprint@\@tempb}{\@tempb:\@tempc}{\expandafter \expandafter \csname
  mn@eprint@\@tempb\endcsname \expandafter{\@tempc}}}

\bibitem[\protect\citeauthoryear{{Abramowicz}, {Czerny}, {Lasota}  \&
  {Szuszkiewicz}}{{Abramowicz} et~al.}{1988}]{abramowicz1988}
{Abramowicz} M.~A.,  {Czerny} B.,  {Lasota} J.~P.,   {Szuszkiewicz} E.,  1988,
  \mn@doi [\apj] {10.1086/166683}, \href
  {https://ui.adsabs.harvard.edu/abs/1988ApJ...332..646A} {332, 646}

\bibitem[\protect\citeauthoryear{{Abramowicz}, {Chen}, {Kato}, {Lasota}  \&
  {Regev}}{{Abramowicz} et~al.}{1995}]{abramowicz1995}
{Abramowicz} M.~A.,  {Chen} X.,  {Kato} S.,  {Lasota} J.-P.,   {Regev} O.,
  1995, \mn@doi [\apjl] {10.1086/187709}, \href
  {https://ui.adsabs.harvard.edu/abs/1995ApJ...438L..37A} {438, L37}

\bibitem[\protect\citeauthoryear{{Bai} \& {Stone}}{{Bai} \&
  {Stone}}{2013}]{bai2013}
{Bai} X.-N.,  {Stone} J.~M.,  2013, \mn@doi [\apj]
  {10.1088/0004-637X/769/1/76}, \href
  {https://ui.adsabs.harvard.edu/abs/2013ApJ...769...76B} {769, 76}

\bibitem[\protect\citeauthoryear{{Begelman}}{{Begelman}}{1979}]{begelman1979}
{Begelman} M.~C.,  1979, \mn@doi [\mnras] {10.1093/mnras/187.2.237}, \href
  {https://ui.adsabs.harvard.edu/abs/1979MNRAS.187..237B} {187, 237}

\bibitem[\protect\citeauthoryear{{Begelman} \& {Armitage}}{{Begelman} \&
  {Armitage}}{2014}]{begelman2014}
{Begelman} M.~C.,  {Armitage} P.~J.,  2014, \mn@doi [\apjl]
  {10.1088/2041-8205/782/2/L18}, \href
  {https://ui.adsabs.harvard.edu/abs/2014ApJ...782L..18B} {782, L18}

\bibitem[\protect\citeauthoryear{{Begelman} \& {Meier}}{{Begelman} \&
  {Meier}}{1982}]{begelman1982}
{Begelman} M.~C.,  {Meier} D.~L.,  1982, \mn@doi [\apj] {10.1086/159688}, \href
  {https://ui.adsabs.harvard.edu/abs/1982ApJ...253..873B} {253, 873}

\bibitem[\protect\citeauthoryear{{Begelman}, {Armitage}  \&
  {Reynolds}}{{Begelman} et~al.}{2015}]{begelman2015}
{Begelman} M.~C.,  {Armitage} P.~J.,   {Reynolds} C.~S.,  2015, \mn@doi [\apj]
  {10.1088/0004-637X/809/2/118}, \href
  {https://ui.adsabs.harvard.edu/abs/2015ApJ...809..118B} {809, 118}

\bibitem[\protect\citeauthoryear{{Belloni}}{{Belloni}}{2010}]{belloni2010}
{Belloni} T.~M.,  2010, in {Belloni} T.,  ed., , Vol.~794, Lecture Notes in
  Physics, Berlin Springer Verlag.
p.~53, \mn@doi{10.1007/978-3-540-76937-8\_3}

\bibitem[\protect\citeauthoryear{{Bhuvana}, {Radhika}, {Agrawal}, {Mandal}  \&
  {Nandi}}{{Bhuvana} et~al.}{2021}]{bhuvana2021}
{Bhuvana} G.~R.,  {Radhika} D.,  {Agrawal} V.~K.,  {Mandal} S.,   {Nandi} A.,
  2021, \mn@doi [\mnras] {10.1093/mnras/staa4012}, \href
  {https://ui.adsabs.harvard.edu/abs/2021MNRAS.501.5457B} {501, 5457}

\bibitem[\protect\citeauthoryear{{Bisnovatyi-Kogan} \&
  {Ruzmaikin}}{{Bisnovatyi-Kogan} \& {Ruzmaikin}}{1974}]{bisnovatyi-kogan1974}
{Bisnovatyi-Kogan} G.~S.,  {Ruzmaikin} A.~A.,  1974, \mn@doi [\apss]
  {10.1007/BF00642237}, \href
  {https://ui.adsabs.harvard.edu/abs/1974Ap&SS..28...45B} {28, 45}

\bibitem[\protect\citeauthoryear{{Cao}}{{Cao}}{2016}]{caoxw2016}
{Cao} X.,  2016, \mn@doi [\apj] {10.3847/0004-637X/817/1/71}, \href
  {https://ui.adsabs.harvard.edu/abs/2016ApJ...817...71C} {817, 71}

\bibitem[\protect\citeauthoryear{{Chen} \& {Taam}}{{Chen} \&
  {Taam}}{1993}]{chen1993}
{Chen} X.,  {Taam} R.~E.,  1993, \mn@doi [\apj] {10.1086/172916}, \href
  {https://ui.adsabs.harvard.edu/abs/1993ApJ...412..254C} {412, 254}

\bibitem[\protect\citeauthoryear{{Chen}, {Abramowicz}, {Lasota}, {Narayan}  \&
  {Yi}}{{Chen} et~al.}{1995}]{chen1995}
{Chen} X.,  {Abramowicz} M.~A.,  {Lasota} J.-P.,  {Narayan} R.,   {Yi} I.,
  1995, \mn@doi [\apjl] {10.1086/187836}, \href
  {https://ui.adsabs.harvard.edu/abs/1995ApJ...443L..61C} {443, L61}

\bibitem[\protect\citeauthoryear{{Davis}, {Stone}  \& {Pessah}}{{Davis}
  et~al.}{2010}]{davis2010}
{Davis} S.~W.,  {Stone} J.~M.,   {Pessah} M.~E.,  2010, \mn@doi [\apj]
  {10.1088/0004-637X/713/1/52}, \href
  {https://ui.adsabs.harvard.edu/abs/2010ApJ...713...52D} {713, 52}

\bibitem[\protect\citeauthoryear{{Dexter}, {McKinney}, {Markoff}  \&
  {Tchekhovskoy}}{{Dexter} et~al.}{2014}]{dexter2014}
{Dexter} J.,  {McKinney} J.~C.,  {Markoff} S.,   {Tchekhovskoy} A.,  2014,
  \mn@doi [\mnras] {10.1093/mnras/stu581}, \href
  {https://ui.adsabs.harvard.edu/abs/2014MNRAS.440.2185D} {440, 2185}

\bibitem[\protect\citeauthoryear{{Done}, {Gierli{\'n}ski}  \& {Kubota}}{{Done}
  et~al.}{2007}]{done2007}
{Done} C.,  {Gierli{\'n}ski} M.,   {Kubota} A.,  2007, \mn@doi [\aapr]
  {10.1007/s00159-007-0006-1}, \href
  {https://ui.adsabs.harvard.edu/abs/2007A&ARv..15....1D} {15, 1}

\bibitem[\protect\citeauthoryear{{Dubus}, {Hameury}  \& {Lasota}}{{Dubus}
  et~al.}{2001}]{dubus2001}
{Dubus} G.,  {Hameury} J.~M.,   {Lasota} J.~P.,  2001, \mn@doi [\aap]
  {10.1051/0004-6361:20010632}, \href
  {https://ui.adsabs.harvard.edu/abs/2001A&A...373..251D} {373, 251}

\bibitem[\protect\citeauthoryear{{Dunn}, {Fender}, {K{\"o}rding}, {Belloni}  \&
  {Cabanac}}{{Dunn} et~al.}{2010}]{dunn2010}
{Dunn} R.~J.~H.,  {Fender} R.~P.,  {K{\"o}rding} E.~G.,  {Belloni} T.,
  {Cabanac} C.,  2010, \mn@doi [\mnras] {10.1111/j.1365-2966.2010.16114.x},
  \href {https://ui.adsabs.harvard.edu/abs/2010MNRAS.403...61D} {403, 61}

\bibitem[\protect\citeauthoryear{{Esin}, {McClintock}  \& {Narayan}}{{Esin}
  et~al.}{1997}]{esin1997}
{Esin} A.~A.,  {McClintock} J.~E.,   {Narayan} R.,  1997, \mn@doi [\apj]
  {10.1086/304829}, \href
  {https://ui.adsabs.harvard.edu/abs/1997ApJ...489..865E} {489, 865}

\bibitem[\protect\citeauthoryear{{Fender} et~al.,}{{Fender}
  et~al.}{1999}]{fender1999}
{Fender} R.,  et~al., 1999, \mn@doi [\apjl] {10.1086/312128}, \href
  {https://ui.adsabs.harvard.edu/abs/1999ApJ...519L.165F} {519, L165}

\bibitem[\protect\citeauthoryear{{Gammie}}{{Gammie}}{2001}]{gammie2001}
{Gammie} C.~F.,  2001, \mn@doi [\apj] {10.1086/320631}, \href
  {https://ui.adsabs.harvard.edu/abs/2001ApJ...553..174G} {553, 174}

\bibitem[\protect\citeauthoryear{{Hameury}}{{Hameury}}{2020}]{hameury2020}
{Hameury} J.~M.,  2020, \mn@doi [Advances in Space Research]
  {10.1016/j.asr.2019.10.022}, \href
  {https://ui.adsabs.harvard.edu/abs/2020AdSpR..66.1004H} {66, 1004}

\bibitem[\protect\citeauthoryear{{Hartmann}, {Calvet}, {Gullbring}  \&
  {D'Alessio}}{{Hartmann} et~al.}{1998}]{hartmann1998}
{Hartmann} L.,  {Calvet} N.,  {Gullbring} E.,   {D'Alessio} P.,  1998, \mn@doi
  [\apj] {10.1086/305277}, \href
  {https://ui.adsabs.harvard.edu/abs/1998ApJ...495..385H} {495, 385}

\bibitem[\protect\citeauthoryear{{Homan}, {Wijnands}, {van der Klis},
  {Belloni}, {van Paradijs}, {Klein-Wolt}, {Fender}  \& {M{\'e}ndez}}{{Homan}
  et~al.}{2001}]{homan2001}
{Homan} J.,  {Wijnands} R.,  {van der Klis} M.,  {Belloni} T.,  {van Paradijs}
  J.,  {Klein-Wolt} M.,  {Fender} R.,   {M{\'e}ndez} M.,  2001, \mn@doi [\apjs]
  {10.1086/318954}, \href
  {https://ui.adsabs.harvard.edu/abs/2001ApJS..132..377H} {132, 377}

\bibitem[\protect\citeauthoryear{{Ichimaru}}{{Ichimaru}}{1977}]{ichimaru1977}
{Ichimaru} S.,  1977, \mn@doi [\apj] {10.1086/155314}, \href
  {https://ui.adsabs.harvard.edu/abs/1977ApJ...214..840I} {214, 840}

\bibitem[\protect\citeauthoryear{{Katz}, {Kolodny}  \& {Nissenbaum}}{{Katz}
  et~al.}{1977}]{katz1977}
{Katz} A.,  {Kolodny} Y.,   {Nissenbaum} A.,  1977, \mn@doi [\gca]
  {10.1016/0016-7037(77)90172-7}, \href
  {https://ui.adsabs.harvard.edu/abs/1977GeCoA..41.1609K} {41, 1609,1613}

\bibitem[\protect\citeauthoryear{{Kempski}, {Quataert}, {Squire}  \&
  {Kunz}}{{Kempski} et~al.}{2019}]{kempski2019}
{Kempski} P.,  {Quataert} E.,  {Squire} J.,   {Kunz} M.~W.,  2019, \mn@doi
  [\mnras] {10.1093/mnras/stz1111}, \href
  {https://ui.adsabs.harvard.edu/abs/2019MNRAS.486.4013K} {486, 4013}

\bibitem[\protect\citeauthoryear{{King}, {Pringle}  \& {Livio}}{{King}
  et~al.}{2007}]{king2007}
{King} A.~R.,  {Pringle} J.~E.,   {Livio} M.,  2007, \mn@doi [\mnras]
  {10.1111/j.1365-2966.2007.11556.x}, \href
  {https://ui.adsabs.harvard.edu/abs/2007MNRAS.376.1740K} {376, 1740}

\bibitem[\protect\citeauthoryear{{Latter} \& {Papaloizou}}{{Latter} \&
  {Papaloizou}}{2012}]{latter2012}
{Latter} H.~N.,  {Papaloizou} J.~C.~B.,  2012, \mn@doi [\mnras]
  {10.1111/j.1365-2966.2012.21748.x}, \href
  {https://ui.adsabs.harvard.edu/abs/2012MNRAS.426.1107L} {426, 1107}

\bibitem[\protect\citeauthoryear{{Liu} \& {Qiao}}{{Liu} \&
  {Qiao}}{2022}]{liu2022}
{Liu} B.~F.,  {Qiao} E.,  2022, \mn@doi [iScience]
  {10.1016/j.isci.2021.103544}, \href
  {https://ui.adsabs.harvard.edu/abs/2022iSci...25j3544L} {25, 103544}

\bibitem[\protect\citeauthoryear{{Liu}, {Mineshige}, {Meyer},
  {Meyer-Hofmeister}  \& {Kawaguchi}}{{Liu} et~al.}{2002}]{liu2002}
{Liu} B.~F.,  {Mineshige} S.,  {Meyer} F.,  {Meyer-Hofmeister} E.,
  {Kawaguchi} T.,  2002, \mn@doi [\apj] {10.1086/341138}, \href
  {https://ui.adsabs.harvard.edu/abs/2002ApJ...575..117L} {575, 117}

\bibitem[\protect\citeauthoryear{{Liu}, {Meyer}  \& {Meyer-Hofmeister}}{{Liu}
  et~al.}{2005}]{liubf2005}
{Liu} B.~F.,  {Meyer} F.,   {Meyer-Hofmeister} E.,  2005, \mn@doi [\aap]
  {10.1051/0004-6361:20053207}, \href
  {https://ui.adsabs.harvard.edu/abs/2005A&A...442..555L} {442, 555}

\bibitem[\protect\citeauthoryear{{Liu}, {van Paradijs}  \& {van den
  Heuvel}}{{Liu} et~al.}{2007}]{liu2007}
{Liu} Q.~Z.,  {van Paradijs} J.,   {van den Heuvel} E.~P.~J.,  2007, \mn@doi
  [\aap] {10.1051/0004-6361:20077303}, \href
  {https://ui.adsabs.harvard.edu/abs/2007A&A...469..807L} {469, 807}

\bibitem[\protect\citeauthoryear{{Liu}, {Done}  \& {Taam}}{{Liu}
  et~al.}{2011}]{liu2011}
{Liu} B.~F.,  {Done} C.,   {Taam} R.~E.,  2011, \mn@doi [\apj]
  {10.1088/0004-637X/726/1/10}, \href
  {https://ui.adsabs.harvard.edu/abs/2011ApJ...726...10L} {726, 10}

\bibitem[\protect\citeauthoryear{{Liu}, {Taam}, {Qiao}  \& {Yuan}}{{Liu}
  et~al.}{2015}]{liu2015}
{Liu} B.~F.,  {Taam} R.~E.,  {Qiao} E.,   {Yuan} W.,  2015, \mn@doi [\apj]
  {10.1088/0004-637X/806/2/223}, \href
  {https://ui.adsabs.harvard.edu/abs/2015ApJ...806..223L} {806, 223}

\bibitem[\protect\citeauthoryear{{Liu}, {Taam}, {Qiao}  \& {Yuan}}{{Liu}
  et~al.}{2017}]{liu2017}
{Liu} B.~F.,  {Taam} R.~E.,  {Qiao} E.,   {Yuan} W.,  2017, \mn@doi [\apj]
  {10.3847/1538-4357/aa894c}, \href
  {https://ui.adsabs.harvard.edu/abs/2017ApJ...847...96L} {847, 96}

\bibitem[\protect\citeauthoryear{{Maccarone}}{{Maccarone}}{2003}]{maccarone2003a}
{Maccarone} T.~J.,  2003, \mn@doi [\aap] {10.1051/0004-6361:20031146}, \href
  {https://ui.adsabs.harvard.edu/abs/2003A&A...409..697M} {409, 697}

\bibitem[\protect\citeauthoryear{{Maccarone} \& {Coppi}}{{Maccarone} \&
  {Coppi}}{2003}]{maccarone2003b}
{Maccarone} T.~J.,  {Coppi} P.~S.,  2003, \mn@doi [\mnras]
  {10.1046/j.1365-8711.2003.06040.x}, \href
  {https://ui.adsabs.harvard.edu/abs/2003MNRAS.338..189M} {338, 189}

\bibitem[\protect\citeauthoryear{{Mahadevan}}{{Mahadevan}}{1997}]{mahadevan1997}
{Mahadevan} R.,  1997, \mn@doi [\apj] {10.1086/303727}, \href
  {https://ui.adsabs.harvard.edu/abs/1997ApJ...477..585M} {477, 585}

\bibitem[\protect\citeauthoryear{{Manmoto}, {Mineshige}  \&
  {Kusunose}}{{Manmoto} et~al.}{1997}]{manmoto1997}
{Manmoto} T.,  {Mineshige} S.,   {Kusunose} M.,  1997, \mn@doi [\apj]
  {10.1086/304817}, \href
  {https://ui.adsabs.harvard.edu/abs/1997ApJ...489..791M} {489, 791}

\bibitem[\protect\citeauthoryear{{McKinney}, {Tchekhovskoy}  \&
  {Blandford}}{{McKinney} et~al.}{2012}]{mckinney2012}
{McKinney} J.~C.,  {Tchekhovskoy} A.,   {Blandford} R.~D.,  2012, \mn@doi
  [\mnras] {10.1111/j.1365-2966.2012.21074.x}, \href
  {https://ui.adsabs.harvard.edu/abs/2012MNRAS.423.3083M} {423, 3083}

\bibitem[\protect\citeauthoryear{{Meyer-Hofmeister}, {Liu}  \&
  {Meyer}}{{Meyer-Hofmeister} et~al.}{2005}]{meyer2005}
{Meyer-Hofmeister} E.,  {Liu} B.~F.,   {Meyer} F.,  2005, \mn@doi [\aap]
  {10.1051/0004-6361:20041631}, \href
  {https://ui.adsabs.harvard.edu/abs/2005A&A...432..181M} {432, 181}

\bibitem[\protect\citeauthoryear{{Meyer}, {Liu}  \& {Meyer-Hofmeister}}{{Meyer}
  et~al.}{2000a}]{meyer2000b}
{Meyer} F.,  {Liu} B.~F.,   {Meyer-Hofmeister} E.,  2000a, \aap, \href
  {https://ui.adsabs.harvard.edu/abs/2000A&A...354L..67M} {354, L67}

\bibitem[\protect\citeauthoryear{{Meyer}, {Liu}  \& {Meyer-Hofmeister}}{{Meyer}
  et~al.}{2000b}]{meyer2000a}
{Meyer} F.,  {Liu} B.~F.,   {Meyer-Hofmeister} E.,  2000b, \aap, \href
  {https://ui.adsabs.harvard.edu/abs/2000A&A...361..175M} {361, 175}

\bibitem[\protect\citeauthoryear{{Meyer}, {Liu}  \& {Meyer-Hofmeister}}{{Meyer}
  et~al.}{2007}]{meyer2007}
{Meyer} F.,  {Liu} B.~F.,   {Meyer-Hofmeister} E.,  2007, \mn@doi [\aap]
  {10.1051/0004-6361:20066203}, \href
  {https://ui.adsabs.harvard.edu/abs/2007A&A...463....1M} {463, 1}

\bibitem[\protect\citeauthoryear{{Miyamoto}, {Kitamoto}, {Hayashida}  \&
  {Egoshi}}{{Miyamoto} et~al.}{1995}]{miyamoto1995}
{Miyamoto} S.,  {Kitamoto} S.,  {Hayashida} K.,   {Egoshi} W.,  1995, \mn@doi
  [\apjl] {10.1086/187804}, \href
  {https://ui.adsabs.harvard.edu/abs/1995ApJ...442L..13M} {442, L13}

\bibitem[\protect\citeauthoryear{{Narayan}}{{Narayan}}{1996}]{narayan1996}
{Narayan} R.,  1996, \mn@doi [\apj] {10.1086/177136}, \href
  {https://ui.adsabs.harvard.edu/abs/1996ApJ...462..136N} {462, 136}

\bibitem[\protect\citeauthoryear{{Narayan} \& {Yi}}{{Narayan} \&
  {Yi}}{1994}]{narayan1994}
{Narayan} R.,  {Yi} I.,  1994, \mn@doi [\apjl] {10.1086/187381}, \href
  {https://ui.adsabs.harvard.edu/abs/1994ApJ...428L..13N} {428, L13}

\bibitem[\protect\citeauthoryear{{Narayan} \& {Yi}}{{Narayan} \&
  {Yi}}{1995a}]{narayan1995a}
{Narayan} R.,  {Yi} I.,  1995a, \mn@doi [\apj] {10.1086/175599}, \href
  {https://ui.adsabs.harvard.edu/abs/1995ApJ...444..231N} {444, 231}

\bibitem[\protect\citeauthoryear{{Narayan} \& {Yi}}{{Narayan} \&
  {Yi}}{1995b}]{narayan1995b}
{Narayan} R.,  {Yi} I.,  1995b, \mn@doi [\apj] {10.1086/176343}, \href
  {https://ui.adsabs.harvard.edu/abs/1995ApJ...452..710N} {452, 710}

\bibitem[\protect\citeauthoryear{{Narayan}, {Igumenshchev}  \&
  {Abramowicz}}{{Narayan} et~al.}{2003}]{narayan2003}
{Narayan} R.,  {Igumenshchev} I.~V.,   {Abramowicz} M.~A.,  2003, \mn@doi
  [\pasj] {10.1093/pasj/55.6.L69}, \href
  {https://ui.adsabs.harvard.edu/abs/2003PASJ...55L..69N} {55, L69}

\bibitem[\protect\citeauthoryear{{Nowak}, {Wilms}  \& {Dove}}{{Nowak}
  et~al.}{2002}]{nowak2002}
{Nowak} M.~A.,  {Wilms} J.,   {Dove} J.~B.,  2002, \mn@doi [\mnras]
  {10.1046/j.1365-8711.2002.05353.x}, \href
  {https://ui.adsabs.harvard.edu/abs/2002MNRAS.332..856N} {332, 856}

\bibitem[\protect\citeauthoryear{{Ohsuga}, {Mori}, {Nakamoto}  \&
  {Mineshige}}{{Ohsuga} et~al.}{2005}]{ohsuga2005}
{Ohsuga} K.,  {Mori} M.,  {Nakamoto} T.,   {Mineshige} S.,  2005, \mn@doi
  [\apj] {10.1086/430728}, \href
  {https://ui.adsabs.harvard.edu/abs/2005ApJ...628..368O} {628, 368}

\bibitem[\protect\citeauthoryear{{Petrucci}, {Ferreira}, {Henri}  \&
  {Pelletier}}{{Petrucci} et~al.}{2008}]{petrucci2008}
{Petrucci} P.-O.,  {Ferreira} J.,  {Henri} G.,   {Pelletier} G.,  2008, \mn@doi
  [\mnras] {10.1111/j.1745-3933.2008.00439.x}, \href
  {https://ui.adsabs.harvard.edu/abs/2008MNRAS.385L..88P} {385, L88}

\bibitem[\protect\citeauthoryear{{Pringle}}{{Pringle}}{1981}]{pringle1981}
{Pringle} J.~E.,  1981, \mn@doi [\araa] {10.1146/annurev.aa.19.090181.001033},
  \href {https://ui.adsabs.harvard.edu/abs/1981ARA&A..19..137P} {19, 137}

\bibitem[\protect\citeauthoryear{{Qiao} \& {Liu}}{{Qiao} \&
  {Liu}}{2009}]{qiao2009}
{Qiao} E.,  {Liu} B.~F.,  2009, \mn@doi [\pasj] {10.1093/pasj/61.2.403}, \href
  {https://ui.adsabs.harvard.edu/abs/2009PASJ...61..403Q} {61, 403}

\bibitem[\protect\citeauthoryear{{Qiao} \& {Liu}}{{Qiao} \&
  {Liu}}{2013}]{qiao2013}
{Qiao} E.,  {Liu} B.~F.,  2013, \mn@doi [\apj] {10.1088/0004-637X/764/1/2},
  \href {https://ui.adsabs.harvard.edu/abs/2013ApJ...764....2Q} {764, 2}

\bibitem[\protect\citeauthoryear{{Qiao} \& {Liu}}{{Qiao} \&
  {Liu}}{2017}]{qiao2017}
{Qiao} E.,  {Liu} B.~F.,  2017, \mn@doi [\mnras] {10.1093/mnras/stx121}, \href
  {https://ui.adsabs.harvard.edu/abs/2017MNRAS.467..898Q} {467, 898}

\bibitem[\protect\citeauthoryear{{Qiao} \& {Liu}}{{Qiao} \&
  {Liu}}{2018}]{qiao2018}
{Qiao} E.,  {Liu} B.~F.,  2018, \mn@doi [\mnras] {10.1093/mnras/sty652}, \href
  {https://ui.adsabs.harvard.edu/abs/2018MNRAS.477..210Q} {477, 210}

\bibitem[\protect\citeauthoryear{{Remillard} \& {McClintock}}{{Remillard} \&
  {McClintock}}{2006}]{remillard2006}
{Remillard} R.~A.,  {McClintock} J.~E.,  2006, \mn@doi [\araa]
  {10.1146/annurev.astro.44.051905.092532}, \href
  {https://ui.adsabs.harvard.edu/abs/2006ARA&A..44...49R} {44, 49}

\bibitem[\protect\citeauthoryear{{Rice}, {Lodato}  \& {Armitage}}{{Rice}
  et~al.}{2005}]{rice2005}
{Rice} W.~K.~M.,  {Lodato} G.,   {Armitage} P.~J.,  2005, \mn@doi [\mnras]
  {10.1111/j.1745-3933.2005.00105.x}, \href
  {https://ui.adsabs.harvard.edu/abs/2005MNRAS.364L..56R} {364, L56}

\bibitem[\protect\citeauthoryear{{Salvesen}, {Simon}, {Armitage}  \&
  {Begelman}}{{Salvesen} et~al.}{2016}]{salvesen2016}
{Salvesen} G.,  {Simon} J.~B.,  {Armitage} P.~J.,   {Begelman} M.~C.,  2016,
  \mn@doi [\mnras] {10.1093/mnras/stw029}, \href
  {https://ui.adsabs.harvard.edu/abs/2016MNRAS.457..857S} {457, 857}

\bibitem[\protect\citeauthoryear{{Shakura} \& {Sunyaev}}{{Shakura} \&
  {Sunyaev}}{1973}]{shakura1973}
{Shakura} N.~I.,  {Sunyaev} R.~A.,  1973, \aap, \href
  {https://ui.adsabs.harvard.edu/abs/1973A&A....24..337S} {24, 337}

\bibitem[\protect\citeauthoryear{{Shapiro}, {Lightman}  \& {Eardley}}{{Shapiro}
  et~al.}{1976}]{shapiro1973}
{Shapiro} S.~L.,  {Lightman} A.~P.,   {Eardley} D.~M.,  1976, \mn@doi [\apj]
  {10.1086/154162}, \href
  {https://ui.adsabs.harvard.edu/abs/1976ApJ...204..187S} {204, 187}

\bibitem[\protect\citeauthoryear{{Simon}, {Beckwith}  \& {Armitage}}{{Simon}
  et~al.}{2012}]{simon2012}
{Simon} J.~B.,  {Beckwith} K.,   {Armitage} P.~J.,  2012, \mn@doi [\mnras]
  {10.1111/j.1365-2966.2012.20835.x}, \href
  {https://ui.adsabs.harvard.edu/abs/2012MNRAS.422.2685S} {422, 2685}

\bibitem[\protect\citeauthoryear{{Smak}}{{Smak}}{1999}]{smak1999}
{Smak} J.,  1999, \actaa, \href
  {https://ui.adsabs.harvard.edu/abs/1999AcA....49..391S} {49, 391}

\bibitem[\protect\citeauthoryear{{Stepney} \& {Guilbert}}{{Stepney} \&
  {Guilbert}}{1983}]{stepney1983}
{Stepney} S.,  {Guilbert} P.~W.,  1983, \mn@doi [\mnras]
  {10.1093/mnras/204.4.1269}, \href
  {https://ui.adsabs.harvard.edu/abs/1983MNRAS.204.1269S} {204, 1269}

\bibitem[\protect\citeauthoryear{{Taam}, {Liu}, {Yuan}  \& {Qiao}}{{Taam}
  et~al.}{2012}]{taam2012}
{Taam} R.~E.,  {Liu} B.~F.,  {Yuan} W.,   {Qiao} E.,  2012, \mn@doi [\apj]
  {10.1088/0004-637X/759/1/65}, \href
  {https://ui.adsabs.harvard.edu/abs/2012ApJ...759...65T} {759, 65}

\bibitem[\protect\citeauthoryear{{Taam}, {Qiao}, {Liu}  \&
  {Meyer-Hofmeister}}{{Taam} et~al.}{2018}]{taam2018}
{Taam} R.~E.,  {Qiao} E.,  {Liu} B.~F.,   {Meyer-Hofmeister} E.,  2018, \mn@doi
  [\apj] {10.3847/1538-4357/aac50d}, \href
  {https://ui.adsabs.harvard.edu/abs/2018ApJ...860..166T} {860, 166}

\bibitem[\protect\citeauthoryear{{Tanaka} \& {Shibazaki}}{{Tanaka} \&
  {Shibazaki}}{1996}]{tanaka1996}
{Tanaka} Y.,  {Shibazaki} N.,  1996, \mn@doi [\araa]
  {10.1146/annurev.astro.34.1.607}, \href
  {https://ui.adsabs.harvard.edu/abs/1996ARA&A..34..607T} {34, 607}

\bibitem[\protect\citeauthoryear{{Tananbaum}, {Gursky}, {Kellogg}, {Giacconi}
  \& {Jones}}{{Tananbaum} et~al.}{1972}]{tananbaum1972}
{Tananbaum} H.,  {Gursky} H.,  {Kellogg} E.,  {Giacconi} R.,   {Jones} C.,
  1972, \mn@doi [\apjl] {10.1086/181042}, \href
  {https://ui.adsabs.harvard.edu/abs/1972ApJ...177L...5T} {177, L5}

\bibitem[\protect\citeauthoryear{{Tetarenko}, {Lasota}, {Heinke}, {Dubus}  \&
  {Sivakoff}}{{Tetarenko} et~al.}{2018}]{tetarenko2018}
{Tetarenko} B.~E.,  {Lasota} J.~P.,  {Heinke} C.~O.,  {Dubus} G.,   {Sivakoff}
  G.~R.,  2018, \mn@doi [\nat] {10.1038/nature25159}, \href
  {https://ui.adsabs.harvard.edu/abs/2018Natur.554...69T} {554, 69}

\bibitem[\protect\citeauthoryear{{Vahdat Motlagh}, {Kalemci}  \&
  {Maccarone}}{{Vahdat Motlagh} et~al.}{2019}]{motlagh2019}
{Vahdat Motlagh} A.,  {Kalemci} E.,   {Maccarone} T.~J.,  2019, \mn@doi
  [\mnras] {10.1093/mnras/stz569}, \href
  {https://ui.adsabs.harvard.edu/abs/2019MNRAS.485.2744V} {485, 2744}

\bibitem[\protect\citeauthoryear{{Xie} \& {Yuan}}{{Xie} \&
  {Yuan}}{2012}]{xie2012}
{Xie} F.-G.,  {Yuan} F.,  2012, \mn@doi [\mnras]
  {10.1111/j.1365-2966.2012.22030.x}, \href
  {https://ui.adsabs.harvard.edu/abs/2012MNRAS.427.1580X} {427, 1580}

\bibitem[\protect\citeauthoryear{{Yan}, {Zhang}  \& {Zhang}}{{Yan}
  et~al.}{2015}]{yan2015}
{Yan} D.,  {Zhang} L.,   {Zhang} S.-N.,  2015, \mn@doi [\mnras]
  {10.1093/mnras/stv2091}, \href
  {https://ui.adsabs.harvard.edu/abs/2015MNRAS.454.1310Y} {454, 1310}

\bibitem[\protect\citeauthoryear{{Yu} \& {Yan}}{{Yu} \& {Yan}}{2009}]{yu2009}
{Yu} W.,  {Yan} Z.,  2009, \mn@doi [\apj] {10.1088/0004-637X/701/2/1940}, \href
  {https://ui.adsabs.harvard.edu/abs/2009ApJ...701.1940Y} {701, 1940}

\bibitem[\protect\citeauthoryear{{Yu}, {van der Klis}  \& {Fender}}{{Yu}
  et~al.}{2004}]{yu2004}
{Yu} W.,  {van der Klis} M.,   {Fender} R.,  2004, \mn@doi [\apjl]
  {10.1086/423953}, \href
  {https://ui.adsabs.harvard.edu/abs/2004ApJ...611L.121Y} {611, L121}

\bibitem[\protect\citeauthoryear{{Yu}, {Lamb}, {Fender}  \& {van der
  Klis}}{{Yu} et~al.}{2007}]{yu2007}
{Yu} W.,  {Lamb} F.~K.,  {Fender} R.,   {van der Klis} M.,  2007, \mn@doi
  [\apj] {10.1086/518733}, \href
  {https://ui.adsabs.harvard.edu/abs/2007ApJ...663.1309Y} {663, 1309}

\bibitem[\protect\citeauthoryear{{Yuan} \& {Narayan}}{{Yuan} \&
  {Narayan}}{2014}]{yuan2014}
{Yuan} F.,  {Narayan} R.,  2014, \mn@doi [\araa]
  {10.1146/annurev-astro-082812-141003}, \href
  {https://ui.adsabs.harvard.edu/abs/2014ARA&A..52..529Y} {52, 529}

\bibitem[\protect\citeauthoryear{{Zdziarski}, {Gierli{\'n}ski},
  {Miko{\l}ajewska}, {Wardzi{\'n}ski}, {Smith}, {Harmon}  \&
  {Kitamoto}}{{Zdziarski} et~al.}{2004}]{zdziarski2004}
{Zdziarski} A.~A.,  {Gierli{\'n}ski} M.,  {Miko{\l}ajewska} J.,
  {Wardzi{\'n}ski} G.,  {Smith} D.~M.,  {Harmon} B.~A.,   {Kitamoto} S.,  2004,
  \mn@doi [\mnras] {10.1111/j.1365-2966.2004.07830.x}, \href
  {https://ui.adsabs.harvard.edu/abs/2004MNRAS.351..791Z} {351, 791}

\makeatother
\end{thebibliography}

\bsp	
\label{lastpage}
\end{document}